\documentclass[reprint,amsmath,amssymb,aps,pra,floatfix]{revtex4-2}
\usepackage{xcolor}
\usepackage{amsmath}
\usepackage{float}
\usepackage{graphics}
\usepackage{dcolumn}
\usepackage{bm}
\usepackage{hyperref}

\usepackage{epsfig}
\usepackage{subfigure}
\usepackage{graphicx}
\usepackage{dcolumn}
\usepackage{stmaryrd}
\usepackage{mathrsfs}
\usepackage{pifont}
\usepackage{amsthm}
\usepackage{amssymb}
\usepackage{mathtools}
\usepackage{bm}
\usepackage{latexsym}
\usepackage{color}

\newcommand{\red}[1]{\textcolor{red}{#1}}

\def\pra#1{{ Phys.\ Rev. A\/} {\bf#1}}

\def\pre#1{{ Phys.\ Rev. E\/} {\bf#1}}
\def\prl#1{{ Phys.\ Rev.\ Lett.} {\bf#1}}

\def\sci#1{{ Science} {\bf#1}}

\def\pla#1{{ Phys.\ Lett. A\/} {\bf#1}}

\def\nat#1{{ Nature} {\bf#1}}

\def\njp#1{{ New. J. \ Phys.} {\bf#1}}

\begin{document}

\preprint{APS/123-QED}

\title{Quantum Next-Generation Reservoir Computing and Its Quantum Optical Implementation }%

\author{Longhan Wang}
 \altaffiliation{These authors contributed equally to this work.}
 
\author{Peijie Sun}
 \altaffiliation{These authors contributed equally to this work.}
  
\author{Ling-Jun Kong}
  
\author{Yifan Sun}
 \email{Contact author: yfsun@bit.edu.cn}
 
\author{Xiangdong Zhang}
 \email{Contact author: zhangxd@bit.edu.cn}

\affiliation {Key Laboratory of advanced optoelectronic quantum architecture and measurements of Ministry of Education}
\affiliation {Beijing Key Laboratory of Nanophotonics \& Ultrafine Optoelectronic Systems, School of Physics, Beijing Institute of Technology, 100081, Beijing, Chinan}

\date{\today}

\begin{abstract}
Quantum reservoir computing (QRC) exploits the information-processing capabilities of quantum systems to tackle time-series forecasting tasks, which is expected to be superior to their classical counterparts. By far, many QRC schemes have been theoretically proposed. However, most of these schemes involves long-time evolution of quantum systems or networks with quantum gates. This poses a challenge for practical implementation of these schemes, as precise manipulation of quantum systems is crucial, and this level of control is currently hard to achieve with the existing state of quantum technology. Here, we propose a different way of QRC scheme, which is friendly to experimental realization. It implements the quantum version of nonlinear vector autoregression, extracting linear and nonlinear features of quantum data by measurements. Thus, the evolution of complex networks of quantum gates can be avoided. Compared to other QRC schemes, our proposal also achieves an advance by effectively reducing the necessary training data for reliable predictions in time-series forecasting tasks. Furthermore, we experimentally verify our proposal by performing the forecasting tasks, and the observation matches well with the theorectial ones. Our work opens up a new way toward complex tasks to be solved by using the QRC, which can herald the next generation of the QRC.
\end{abstract}
\maketitle
\section{\label{sec:1}INTRODUCTION}
Machine learning (ML) has achieved dramatic success over the past decade with applications in computer vision, natural language processing, medical diagnosis, time-dependent signal processing, search engines, etc.\cite{Jordan_2015,bishop2006pattern}. Among various types of ML, reservoir computing \cite{Jaeger_2004,Milano_2021,Pathak_2018,Gauthier_2021,Tanaka_2019,Yan_2022,Cucchi_2021,Zhong_2022,Chen_2023,Tan_2023,Zhong_2021,Vandoorne_2014,Sun_2021,Appeltant_2011,Rafayelyan_2020,Liu_2022,Brunner_2013,Woods_2012,Du_2017,Wei_2022,Moon_2019,Akashi_2022}, or its extension in the differential equation form \cite{domingo2023anticipating}, has received a lot of attention as a special recurrent neural network (RNN), because it is a best-in-class and easily implementable strategy for processing information generated by dynamical system, via analyzing the observed time-series data. 

In contrast to the classical reservoir computing, the quantum reservoir computing (QRC) \cite{Nakajima_2019,Spagnolo_2022,Fujii_2017,tran2020higher,Mart_nez_Pe_a_2020,Suzuki_2022,Ghosh_2021,martinez2023quantum,Bravo_2022,Govia_2021,Ghosh_2019,Gross_2010,ghosh2021realising,Angelatos_2021,nakajima2018reservoir,Garc_a_Beni_2023,Chen_2020,Agnew_2011,Llodr__2022,Nokkala_2024,Sornsaeng_2024} is expected to be superior to their classical counterparts. By far, series of QRC schemes have been proposed, which are used for image classification by sequential data analysis \cite{Spagnolo_2022}, time-dependent signal processing \cite{Fujii_2017}, quantum state preparation \cite{Suzuki_2022}, recognition of quantum entanglement \cite{Ghosh_2019}, phase transition analysis \cite{Mart_nez_Pe_a_2021}, waveform classification task \cite{Dudas_2023} and so on. However, these schemes are all based on the quantum dynamical process or networks of quantum gates \cite{Li_2020,Peruzzo_2014,Li_2015,Kandala_2017,Wang_2023,Cong_2019,Wang_2021,Biamonte_2017,Havl_ek_2019,F_rrutter_2024,West_2023,Haug_2023,Huijgen_2024,Senokosov_2024,Govia_2022}, which are not easy to implement considering the current stage of quantum computing platform. There are several strategies for optimizing the circuit of QRC \cite{Domingo_2022,Domingo_2024}, but the obstacle for the implementation is not entirely removed. Hence, the practical application of these schemes is faced with a significant challenge due to the difficulty in achieving precise control over quantum systems, which is essential for implementation. If not controlled properly, the performance of quantum learning strategies could be worse than classical methods. Currently, most of the work in this area is theoretical, and only a few QRC schemes have been experimentally demonstrated using cloud-based superconducting devices \cite{Garc_a_Beni_2023,Hu_2024,suzuki2022natural,Molteni_2023,Kubota_2023,yasuda2023quantum}, indicating limitations for broader applications. 

In this work, we propose an alternate scheme of QRC that is more amenable to experimental implementation, with further improvements achieved by introducing nonlinear vector autoregression. In fact, our work presents a strategy for implementing the function of next generation reservoir computing (NGRC) in quantum computing \cite{Gauthier_2021}. It is known that NGRC has shown great performance in the prediction of chaotic dynamics. Then, how to combine the advance of NGRC and quantum computing for achieving a more powerful scheme is an intriguing question. {\emph{One solution to the question has been given by Ref.\cite{Sornsaeng_2024}, which investigates many-body quantum dynamics forecasting by a quantum version of NGRC.} Distinguished from the method, we propose to directly prepare the quantum states that encodes the linear and nonlinear feature of the given data. Then, processing of the data can be achieved by only measuring the prepared states and then applying the linear transformation on them, so that deep network of quantum gates or long-time quantum evolution can be avoided. Compared to other QRC schemes, our proposal also displays an advance in data requirements during the training phase. We found that the amount of training data for obtaining a reliable outcome can be effectively decreased by the method in time-series forecasting tasks. Furthermore, we experimentally perform the timer task and Lorenz63, which is a prediction task of a weather system developed by Lorenz in 1963, prediction task by using quantum optical platforms, validating the feasibility of the scheme. Our work opens up a new approach to solving complex tasks using the QRC, which is expected to have a wide range of applications.

\section{\label{sec:2} THE SCHEME OF QUANTUM RESERVOIR COMPUTING WITH NONLINEAR VECTOR AUTOREGRESSION}
To employ the established QRC schemes \cite{Spagnolo_2022,Fujii_2017,Govia_2021,Agnew_2011,Llodr__2022} for tasks like time-series forecasting, it typically requires to input the relevant data into a network of quantum particles or nodes, usually called a quantum reservoir (QR). Then, the QR evolves under its Hamiltonian, and more input data can be fed into it during the evolution. The final output, which may be used as the prediction of a time-series or else, can be obtained by performing linear transformations on the measurements of the QR. In contrast, our QRC scheme replaces the complex quantum network with relatively simpler quantum states that can give the linear and nonlinear features of data. Then, the measurements on the evolution of a quantum network can be replaced by the measurements on the states in our proposal. The fundamental scheme of data processing in our proposal is illustrated in Fig.~\ref{Fig.1}.
\begin{figure}
 \includegraphics{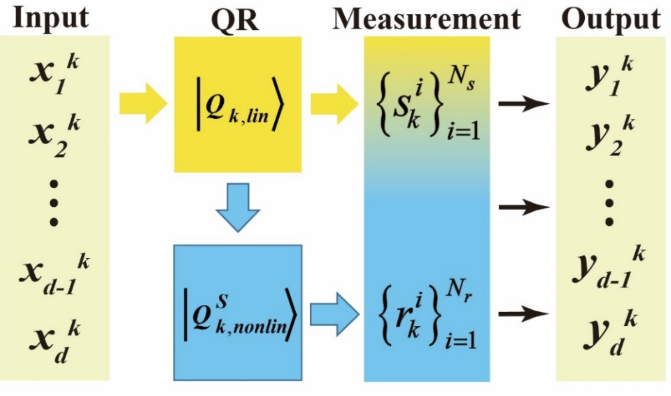}
 \caption{\label{Fig.1}Schematic of QRC with nonlinear vector autoregression. Firstly, based on the input signal $\vec{x}_{k}=\left[x_{1}^{k}, x_{2}^{k} \cdots x_{d}^{k}\right]^{T}$, we obtain state $\left|Q_{k,lin}\right\rangle$. State $\left|Q_{k,nonlin}^S\right\rangle$ is given by the tensor product of $\left|Q_{k,lin}\right\rangle$. Then, projection measurements on both $\left|Q_{k,lin}\right\rangle$ and $\left|Q_{k,nonlin}^S\right\rangle$ are performed. Finally, after applying a linear transformation to the measurement signal, the output data can be given.}
\end{figure}

As shown in Fig.~\ref{Fig.1}, the quantum states that give the linear and nonlinear features of data are respectively represented as $\left|Q_{k,lin}\right\rangle$ and $\left|Q_{k,nonlin}^S\right\rangle$. Suppose that the input signal at the th time step of a discretized dynamical process is given by a $d$-dimensional complex vector $\vec{x}_{k}=\left[x_{1}^{k}, x_{2}^{k} \cdots x_{d}^{k}\right]^{T}$, where represents the matrix transposition. Then, in order to predict the future signals of the process, the amplitudes of the basis states of $\left|Q_{k,lin}\right\rangle$ are set to encode the components of input signal $\vec{x}_k$ and the $(u-1)$ input signals in front of it. By using the computational basis, $\left|Q_{k,lin}\right\rangle$ can be expressed by
\begin{equation}
    \left|Q_{k,lin}\right\rangle=\frac{1}{\sqrt{N}}\left[\vec{x}_{k} \oplus \vec{x}_{k-q} \oplus \vec{x}_{k-2 q} \oplus \cdots \oplus \vec{x}_{k-(u-1) q}\right]^{T},
    \label{Eq.1}
\end{equation}
where $\oplus$ represents the vector concatenation operation. Eq.~(\ref{Eq.1}) indicates that there are $u$ signals encoded by $|Q_{k,lin}\rangle$ in total, with the last signal being $\vec{x}_k$. The real number $N$ is the normalization factor. $q$ is an integer that represents the interval length between the nearest data samples involved in $\left|Q_{k,lin}\right\rangle$ . From Eq.~(\ref{Eq.1}), it can be noticed that  $\left|Q_{k,lin}\right\rangle$ consists of $d\times u$  elements, including the prior data of the time series. According to the general theory of universal approximators \cite{Gonon_2020,Franz_2006}, to predict future data normally requires a great deal of prior data so that  has to be nearly infinite. However, by using the state defined by Eq.~(\ref{Eq.1}), such a requirement turns out to be unnecessary, and truncating $u$  to a smaller value does not result in significant errors. This can be seen from the theory of reproducing kernel Hilbert space \cite{Paulsen_2016,Bollt_2021}, validated by our numerical investigation in the next section. Following Eq.~(\ref{Eq.1}), the state $\left|Q_{k,nonlin}^S\right\rangle$  providing the non-linear feature can be given by the tensor product of  $\left|Q_{k,lin}\right\rangle$
\begin{equation}
    \left|Q_{k,nonlin}^{S}\right\rangle=\left|Q_{k,lin}\right\rangle \otimes\left|Q_{k,lin}\right\rangle \otimes \ldots \otimes\left|Q_{k,lin}\right\rangle,
\end{equation}
where superscript  $S$ of $\left|Q_{k,nonlin}^{S}\right\rangle$  denotes the number of  $\left|Q_{k,lin}\right\rangle$ in the product. In principle, the $\left|Q_{k,nonlin}^{S}\right\rangle$ preparation of   can be achieved by generating $S$  states in form of $\left|Q_{k,lin}\right\rangle$ . Hence, in general, the measurements on the $\left|Q_{k,lin}\right\rangle$  give the linear feature of the input data, which is the linear combination of its components. Similarly, the measurements on the $\left|Q_{k,nonlin}^{S}\right\rangle$  give the nonlinear feature of the input data, which is the combination of all possible products of the components. This is functionally equivalent to the procedure of processing data with the next-generation of reservoir computing scheme, in which the nonlinear feature of data plays an important role \cite{Havl_ek_2019}.

In order to handle given tasks, we propose to set the measurement on $\left|Q_{k,lin}\right\rangle$ and $\left|Q_{k,nonlin}^{S}\right\rangle$ as follows. For the state $\left|Q_{k,lin}\right\rangle$, we first define its measurement set as $\left\{\hat{\lambda}_i\right\}_{i=1}^{N_s}$, where $\hat{\lambda}_i$ is a projection operator composed of computational basis \cite{Nielsen_2012}, and $N_s=d\times u$. Then, the measurement signal of $\left|Q_{k,lin}\right\rangle$ is defined as the square root of the average value of the projection operators, denoted as $\sqrt{\langle\hat{\lambda}_i \rangle}=\sqrt{tr(\hat{\lambda}_i\left|Q_{k,lin}\right\rangle\langle Q_{k,lin}|)}$. Thus, we obtain linear measurement signals of size $d\times u$, represented as $s_k=\left\{ s_k^i\right\}^{d\times u}_{i=1}$. For the state $\left|Q_{k,nonlin}^{S}\right\rangle$, because it is the tensor product of $S$ states $\left|Q_{k,nonlin}^{S}\right\rangle$, the projection operator on it can be given by the tensor product of $\hat{\lambda}_i$. Thus, a set of $(d\times u)^S$ projection operators can be obtained. Here, we choose the subset that has all distinguished projection operators as the measurement set for $\left|Q_{k,nonlin}^{S}\right\rangle$, and size of it adds up to $N_r=(d\times u)(d\times u+1)(d\times u+2)\cdots(d\times u+S-1)/S!$ . Similarly, the measurement signal for $\left|Q_{k,nonlin}^{S}\right\rangle$ is also represented as the square root of the average value of the projection operators in the measurement set. Thus, we can obtain nonlinear measurement signals of size $N_r$, marked as $r_k=\left\{ r_k^i \right\}^{N_r}_{i=1}$. Finally, we write total outcome as a vector $\vec{M}_{k}=\left[s_{k}^{1}, \ldots,~s_{k}^{d \times u}, r_{k}^{1}, \ldots,~r_{k}^{N_{r}}\right]^{T}$, which can be turned to target outputs by linear transformations. Notice that, in the above setup for measurements, the measurement set $\left\{\hat{\lambda}_i\right\}_{i=1}^{N_s}$ of $\left|Q_{k,lin}\right\rangle$ does not have to be exactly the projectors in computational basis. It can be any of their linearly independent combinations, and the same goes for the measurement set of $\left|Q_{k,nonlin}^{S}\right\rangle$. In fact, because linear signal $s_k$  mainly capture the information of input data, it is relatively easy to compute classically. However, nonlinear signal $r_k$ involves all kinds of products of the input data, which is hard to compute, and is here proposed to obtained by measurements.

Next, we explain how to obtain the linear transformation, i.e., the training process. In this process, the data samples of the training set are encoded by quantum states one-by-one, generating a series of $\left|Q_{k,lin}\right\rangle$ and  $\left|Q_{k,nonlin}^{S}\right\rangle$ with different time step $k$. In order to capture the whole character of the training set, $k$ usually runs over all of the input signal set. Suppose that integer $k$ ranges from 1 to $O$. Then, using the above scheme, $O$ outcome vectors in total can be obtained, denoted by $\left\{\vec{M}_{k}\right\}_{k=1}^{O}$. By rearranging the elements in $\left\{\vec{M}_{k}\right\}_{k=1}^{O}$, the total outcome can be expressed by a matrix $m=\left\{m_{k p}\right\}\left(1 \leq k \leq O, 1 \leq p \leq\left(1+N_{S}+N_{r}\right)\right)$, whose row vectors are transposition of all $\vec{M}_{k}$. Meanwhile, we use  $y=\left\{y_{L k}\right\}(1 \leq L \leq d, 1 \leq k \leq O)$ and $\bar{y}=\left\{\bar{y}_{L k}\right\}(1 \leq L \leq d, 1 \leq k \leq O)$ to represent the data matrices of the QRC output signal and the target signal, both of which have a size of $d \times O$. In a least-square sense, we match the QRC output signal $y=W_{ out } m $ with the target signal $ \bar{y}$, and weight matrix  $W_{ out }$ is the linear transformation as required. The problem can be summarized by solving the following equation
\begin{equation}\label{Eq.3}
    \bar{y}=W_{out}m.
\end{equation}
Based on this, we can obtain $W_{out}$ by minimizing the mean squared error. Under the condition that $m$ has full row rank, it is expressed by 
\begin{equation}\label{Eq.4}
    W_{out}=\bar{y}m^T(mm^T)^{-1},
\end{equation}
where $(\cdot)^{-1}$ represents the inverse of matrix.

A typical character of the above scheme is that one does not need to tango the input signal (or the training data set) with a quantum system when it evolves, as in the previous QRC schemes. In the above scheme, if an input signal is given, the encoding of data samples by $\left|Q_{k,lin}\right\rangle$ and  $\left|Q_{k,nonlin}^{S}\right\rangle$ can be performed independently, and no evolution of any state is demanded. This significantly avoids the influence of initial conditions on the reservoir state. Due to the nonlinearity in a QRC, the initial conditions affect the performance deeply. Therefore, one usually needs to let the system evolve under ancillary input data for a duration to eliminate the influence of initial conditions, and such a procedure is sometimes called the warm-up. Because there is no evolution of states involved in our scheme, the warm-up procedure can be greatly suppressed. This character is particularly helpful in the cases when gathering the training data is consuming. Furthermore, our scheme indicates that the fundamental requirements for a QRC is generating the tensor product states like  $\left|Q_{k,nonlin}^{S}\right\rangle$. This is relatively simpler compared to controlling the QR like multi-partite quantum system or networks of quantum gates, which provides a shortcut for the implementations. Next, we validate the performance of our scheme through several benchmark tasks, and illustrate an experimental verification.

\section{\label{sec:3}EXAMPLES}
We consider several benchmark tasks to demonstrate the performance of this scheme. Our tasks include a timer task in a timing control system \cite{Fujii_2017}, a nonlinear autoregressive moving average (NARMA) task in an input-driven or non-autonomous dynamical system \cite{Appeltant_2011,Fujii_2017,Verstraeten_2007}, and chaotic attractor tasks in autonomous dynamical systems \cite{Jaeger_2004,Gauthier_2021,Fujii_2017,tran2020higher,nakajima2018reservoir,Maass_2002}. The chaotic attractor tasks include the Mackey-Glass (MG) prediction task, a prediction task of a weather system developed by Lorenz in 1963 (Lorenz63), and a prediction task of the dynamics of a double-scroll electronic circuit (DSEC). These cover typical applications targeted by QRC schemes. The main text only presents the results of timer task and the Lorenz63 task, as the supports of the following experiments. The NARMA task, the MG task, and the DSEC task are placed in Appendix~\ref{A},~\ref{B}, and~\ref{C}, respectively. For the ease of treatment, we choose the measurement set merely composed of each computational basis. 
\subsection{Timer task}
We first introduce the timer task. The timer task is a benchmark task that can be used to directly evaluate whether a QRC has an exploitable memory function. The input-output relationship of a timer can be expressed as
\begin{eqnarray}
&&x_k=\Big\{\begin{matrix}
  1& (k\ge k') \\
  0& (otherwise)
\end{matrix},\nonumber
\\
&&y_k=\Big\{\begin{matrix}
  1& (k= k'+\tau_{timer}) \\
  0& (otherwise)
\end{matrix}.
\end{eqnarray}
where $k'$ is the time step that prompts the system to start and $\tau_{timer}$ is the duration of the timer. The goal of this task is to simulate this timer by using the QRC scheme. As shown in Fig.~\ref{Fig.2}, the task consists of 10 time steps. At time step 3, i.e., $k'=3$, the input signal changes from 0 to 1 and the system continues to run for the remaining 7 time steps. The output signal is 1 at time step 9 and 0 for the rest of the time steps, i.e., a timer with a duration of 6 is executed. 

Following the above general scheme, here, the linear and nonlinear quantum states are represented according to the input signal, respectively. According to the above theoretical scheme, the parameters of $\left|Q_{k,lin}\right\rangle$ here are taken as $q=1$  and $u=\tau_{timer}+2$  respectively, proportional to $[x_k,~x_{k-1},~x_{k-2},~x_{k-3},~x_{k-4},~x_{k-5},~x_{k-6},~x_{k-7},~x_{k-8}]^T$. The parameter of $\left|Q_{k,nonlin}^{S}\right\rangle$ here is taken as $S=0$, i.e., $\left|Q_{k,nonlin}^{S}\right\rangle=0$. We then numerically simulated \begin{figure}
    \centering
    \includegraphics{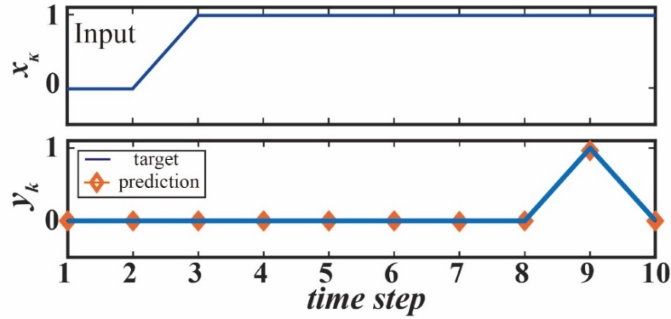}
    \caption{The performance of QRC with nonlinear vector nonlinear vector autoregression in a timer task. The plots show the input signal (upper panel) and output signal (lower panel) as functions of time step in the test phase. The target ones are colored blue and the prediction ones are colored orange.}
    \label{Fig.2}
\end{figure}the timer task and the results of the output variables are shown at the bottom of Fig.~\ref{Fig.2}. The variation of the target signal (blue) and the predicted signal (orange) with time steps for the test phase are plotted separately in the figure. It can be clearly seen that the agreement between the predicted signal and the target signal is very good with $100\%$ accuracy. In addition, to perform the timer task in the traditional QRC scheme \cite{Fujii_2017}, 800 data are required to train the timers with duration $\tau_{timer}=5,~10,~15,~20,~25,$ and 30, respectively. In the above scheme, only $\tau_{timer}+2$ data are required to train the timer with duration $\tau_{timer}$. 

Compared with the conventional QRC, it can be seen that this scheme does not require the provision of warm-up data, thus eliminating the effect of the initial state of the QR on the system, and the amount of training data is largely reduced. Notice that the absence of a warm-up phase in the proposed method does not necessarily imply enhanced data efficiency in training. While achieving the demanded accuracy, there is at least an order of magnitude reduction in the training dataset, even if one get rid of the warm-up data in the previous scheme and only compare the size of the training datasets.
\subsection{Lorenz63 prediction task}
The second typical benchmark task is the Lorenz63 prediction task. The Lorenz63 task is more complex than the timer task, because it is required to predict the dynamics of a chaotic system with a double attractor. Additionally, unlike the previous case, the output of the Lorenz63 system is fed back as the input for the next time step. This means that when the trained reading system generates output, it receives its own output signal as an input signal through feedback connections, rather than obtaining input signals externally. The Lorenz63 consists of three coupled nonlinear differential equations, which can be represented as\begin{eqnarray}
    &&\dot{x}_1=10(x_2-x_1),\nonumber\\
    &&\dot{x}_2=x_1(28-x_3)-x_2,\nonumber\\
    &&\dot{x}_1=x_1x_2-8x_3/3.
    \label{Eq.6}
\end{eqnarray}

To accomplish the Lorenz63 prediction task, we generate 1700 standardized input data samples through numerical integration. The first 200 data samples are used for flushing, followed by 1000 data samples for the training phase and 500 data samples for the testing phase. Next, according to the above theoretical scheme, the parameters of $\left|Q_{k,lin}\right\rangle$ are taken as $q=1$ and $u=2$, proportional to $\left[\vec{x}_{k}\oplus\vec{x}_{k-1}\right]^{T}$. The parameter of $\left|Q_{k,nonlin}^{S}\right\rangle$ is taken as  S=2, i.e., $\left|Q_{k,nonlin}^{2}\right\rangle=\left|Q_{k,lin}\right\rangle \otimes\left|Q_{k,lin}\right\rangle$. Due to the sensitivity of the chaotic model to the data \cite{devaney2018introduction}, after obtaining the measurement results, the normalization factors of the quantum states are multiplied with the measurement results. Then, a linear transformation is performed. The numerical results are shown in Fig. 3, which illustrates the change of $x_{1}^{k}$, $x_{2}^{k}$ and $x_{3}^{k}$ over time steps in the testing phase. The target signal is colored blue, and the predicted signal is colored orange. We compute the normalized root-mean-square (NRMSE) error \cite{Gauthier_2021} for the target signal and the predicted signal for time steps from 1 to 270 in the testing phase. They are 6.3 $\times 10^{-2}$, $7.6 \times 10^{-2}$ and $1.5 \times 10^{-2}$. It is worth mentioning that, in this case, $\left|Q_{k, lin}\right\rangle$ is a 3-qubit state, and $\left|Q_{k,nonlin}^{S}\right\rangle$ is a 6-qubit state. Therefore, one needs at most 6 qubits to perform the prediction of Lorenz63 dynamics by our scheme, which is equivalent to the capacity of the best classical strategy with 28 features \cite{Gauthier_2021}. Furthermore, it can be seen from Fig.~\ref{Fig.3} that our scheme can effectively predict longer dynamics with small error than the classical treatment, by using the same amount of data. This indicates a reduction of the requirements on the amount of training data for the prediction. Actually, the above results can also be verified experimentally, and we demonstrate the scheme in below.\begin{figure}[h]
    \centering
    \includegraphics{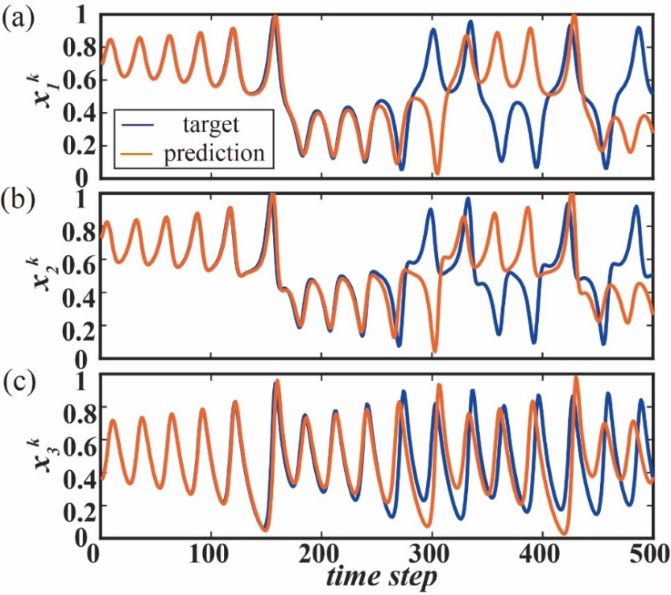}
    \caption{The performance of QRC with nonlinear vector nonlinear vector autoregression in the Lorenz63 prediction task. (a), (b) and (c) are the plots of the dynamics of different components defined by Eq.~(\ref{Eq.6}). The plots show the input signal (upper panel) and output signal (lower panel) as functions of time step in the test phase. The target ones are colored blue and the predicted ones colored orange.}
    \label{Fig.3}
\end{figure}
\section{\label{sec:4}THE EXPERIMENTAL REALIZATION OF THE REALIZATION OF THE QRC WITH NONLINEAR VECTOR AUTOREGRESSION}
In this section, we propose a scheme for implementing QRC with nonlinear vector autoregression, and a schematic diagram is shown in Fig.~\ref{Fig.4}. The setup in Fig.~\ref{Fig.4} performs the functions shown in Fig.~\ref{Fig.1} correspondingly, \begin{figure*}[htbp]
    \includegraphics{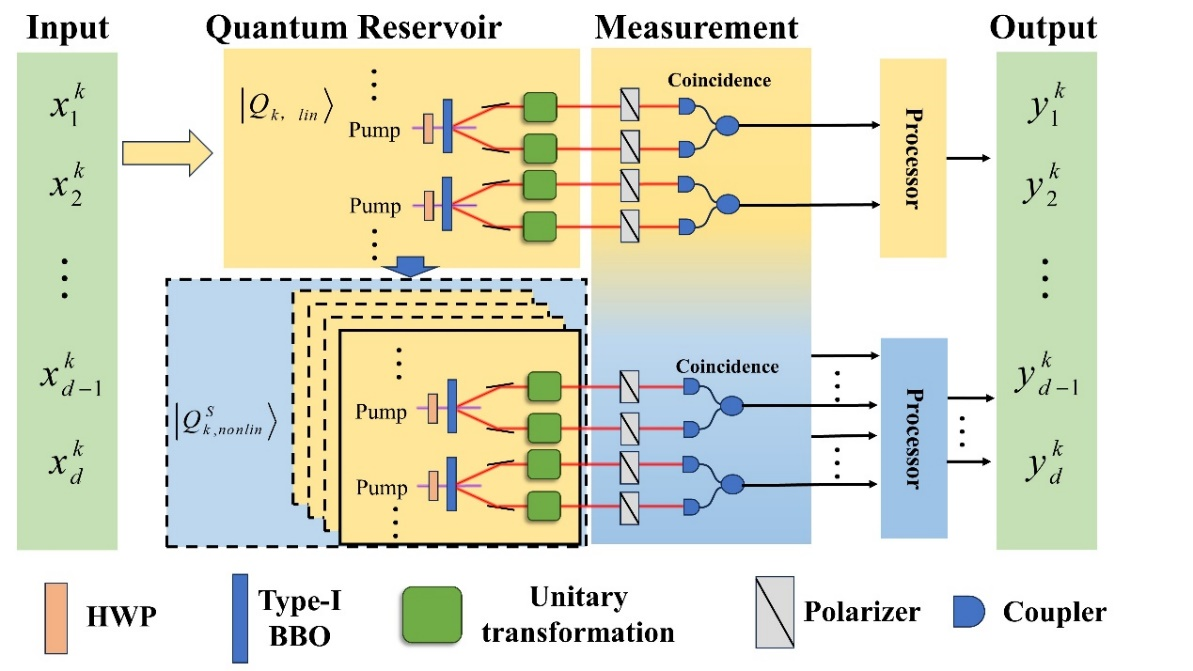}
    \caption{\label{Fig.4}A strategy of implementing the QRC with nonlinear vector autoregression by entangled photons. Firstly, multiple entangled photon pairs are generated by a multi-channel UV pump light source through BBOs, and then the photon pairs are manipulated by unitary transformations. After that, polarizers are utilized for projections on each pair of photons in four directions $,|HH\rangle,~|HV\rangle,~|VH\rangle$ and $|VV\rangle$. Then, the process repeats, and the total signals of the procedure severs as the output of the reservoir. }
\end{figure*}including the preparation and measurement of state   and. As shown in Fig.~\ref{Fig.4}, both $\left|Q_{k,lin}\right\rangle$ and $\left|Q_{k,nonlin}^{S}\right\rangle$ are encoded by multiple entangled photon pairs, each pair of which can be generated by the process of spontaneous parametric down conversion (SPDC). Particularly, the photon state generated by such a setup can be viewed as the composition of the path mode state and the two-photon entangled state, so that the encoded states can be easily generated and manipulated by the local unitary operators. Then, the polarizers are applied for projection operator operation on each photon pair, and coincidence counts of the setup gives the final output signals of the reservoir in the above scheme.

Next, we take the cases involving 3-qubit $\left|Q_{k,lin}\right\rangle$ as examples for experimental illustration, and the schematic is shown in Fig.~\ref{Fig.5}. In our scheme, a pulsed ultraviolet laser, which has 405 nm central wavelength, 140 fs pulse duration, and 80 MHz repetition rate, passes the beam displacer (BD).  The BD is applied for polarizing the light in horizontal ($H$) direction. Subsequently, the beam passes through a half-wave plate (HWP) and a polarized beam splitter (PBS), and is split into two beams, whose polarization state can be denoted by $\alpha H$ and $\beta$, respectively. Here, $V$ represents the vertical direction, and the ratio of  $\alpha$ and $\beta$ is set by the HWP in front of the PBS. These two beams of light, after being manipulated by the quartz plate (QP) and the HWP \cite{Nambu_2002}, pass through two pairs of BBOs, each of which consists of two Type-I BBO with perpendicular optical axes pressed together. This process generates two pairs of entangled photons \cite{Kwiat_1999}, and the state of each photon pair is $\left|\psi\right\rangle=(|HH\rangle+|VV\rangle)/\sqrt{2}$ . The polarization correlation curves of the photon pairs generated by the two BBOs are provided in Appendix \ref{E}, validating their entanglements. By rotating the HWP and QP in front of the two BBOs, we can obtain the entangled photon pairs \cite{Qiang_2018,Wei_2005}, which are expressed as
\begin{eqnarray}
    &&\cos \theta_{1}\left|H_{A_{1}} H_{B_{1}}\right\rangle+e^{i \phi_{1}} \sin \theta_{1}\left|V_{A_{1}} V_{B_{1}}\right\rangle, \label{Eq.7}\\ 
    &&\cos \theta_{2}\left|H_{A_{2}} H_{B_{2}}\right\rangle+e^{i \phi_{2}} \sin \theta_{2}\left|V_{A_{2}} V_{B_{2}}\right\rangle. \label{Eq.8}
\end{eqnarray}
For the phase $e^{i\phi_{1}}$ and $e^{i\phi_{2}}$ in Eq.~(\ref{Eq.7}) and Eq.~(\ref{Eq.8}), we apply QHQs (quarter-wave-plate(QWP), a combination of HWP and QWP) for tuning them precisely. The unitary transformations $\hat{U}$ is then realized by solely HWPs or a combination of HWPs and QWPs for each one of the optical paths. In total, four rotations, denoted by $\hat{U}_{A}$, $\hat{U}_{B}$, $\hat{U}_{A}^{\prime}$, and $\hat{U}_{B}^{\prime}$, are applied in the optical paths, and the details of them are shown in Appendix \ref{F}. Hence, the two pairs of entangled photons can both be arbitrarily tuned. Further, a 3-qubit arbitrarily tunable state can be implemented by the two pairs of photons, described by
\begin{eqnarray}
\left|\varphi\right\rangle=&&\alpha^{\prime}(a|H_{A_{1}}H_{B_{1}}\rangle+b|H_{A_{1}}V_{B_{1}}\rangle+ \nonumber\\\nonumber&&c|V_{A_{1}}H_{B_{1}}\rangle+d|V_{A_{1}}V_{B_{1}}\rangle)+ \\\nonumber&&\beta^{\prime}(a^{\prime}|H_{A_{2}}H_{B_{2}}\rangle+b^{\prime}|H_{A_{2}}V_{B_{2}}\rangle+
\\\nonumber&&c^{\prime}|V_{A_{2}}H_{B_{2}}\rangle+d^{\prime}|V_{A_{2}}V_{B_{2}}\rangle)\\ \nonumber
=&&P_1|H_{A_1}H_{B_1}\rangle+P_2|H_{A_1}V_{B_1}\rangle+\\\nonumber &&P_3|V_{A_1}H_{B_1}\rangle+P_4|V_{A_1}V_{B_1}\rangle+\\\nonumber &&P_5|H_{A_2}H_{B_2}\rangle+P_6|H_{A_2}V_{B_2}\rangle+\\
&&P_7|V_{A_2}H_{B_2}\rangle+P_8|V_{A_2}V_{B_2}\rangle, 
\label{Eq.9}
\end{eqnarray}
where $a,~b,~c$, and $d$ $(a',~b',~c'$, and $d')$ are the coefficients of the basis states determined by rotations $\hat{U} _{A}$ and $\hat{U}_{B}$ ($\hat{U}_{A}'$ and $\hat{U}_{B}'$). $\alpha'$ and $\beta'$ are the total magnitude of the two entangled photon pairs, and they are proportional to $\alpha$ and $\beta$. $P_{1}$ to $P_{8}$ are the products of the corresponding variables. Actually, the “third qubit” in Eq.~(\ref{Eq.9}) is the path qubit state. One benefit of using four photons for implementing a 3-qubit state is that the tuning of the state can be more convenient to perform. As shown by Eq.~(\ref{Eq.9}), the coefficients $P_1$ to $P_8$ can be tuned merely by setting $\hat{U}_A$, $\hat{U}_B$, $\hat{U}_A^{\prime}$, and $\hat{U}_B^{\prime}$, together with $\alpha^{\prime}$ and $\beta^{\prime}$. Based on the above setup, $\alpha^{\prime}$ and $\beta^{\prime}$ can also be tuned by using HWPs (the physical reason is given by Appendix \ref{D}). The relative phase factor between $\alpha'$ and $\beta'$ has no effect on the tasks so that it is not tuned in the experiments.
\begin{figure}[htbp]
    \centering
    \includegraphics[width=3.4in]{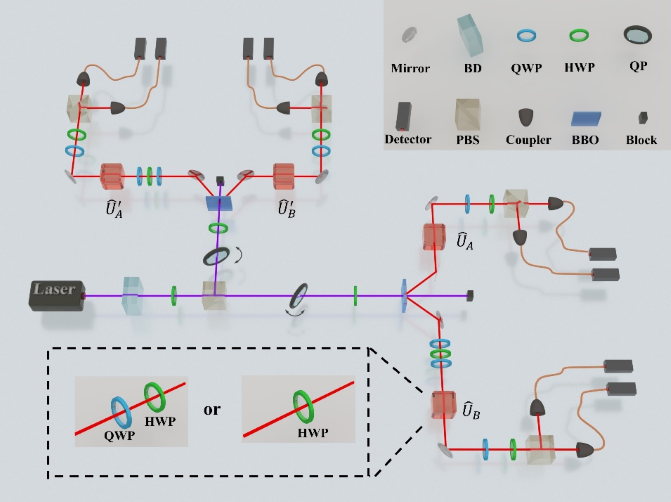}
    \caption{Experimental setup for the preparation of a 3-qubit arbitrary quantum state. An ultrafast ultraviolet pump laser passer BBOs to generate entangled photon pairs. The HWP in front of the first PBS is adjusted to change the light intensity of the two paths, which allows us to modulate the number of photons pumped by the two pairs BBOs. Subsequently, our desired quantum states are mainly prepared by four local rotations $\hat{U}_A,\hat{U}_B,\hat{U}_A^{\prime}$, and $\hat{U}_B^{\prime}$. Each local rotations can be realized by HWP or a combination of HWP and QWP. Finally, projection measurements are made on each entangled photon pair to obtain the coincidence count.}
    \label{Fig.5}
\end{figure}

Then, we discuss the projective measurement operation in the experiment. In this process, two pairs of entangled photons are projected measurement by a device consisting of an HWP, a QWP and a PBS, which implements the polarizer in Fig.~\ref{Fig.4}. The coincidence counts are measured by a pair of avalanche photodiodes (APDs). Further, the result corresponding to the projection measurement can be obtained and expressed as $\left\langle\hat{\lambda}_{R,N}\right\rangle=tr\left(\hat{\lambda}_{R,N}|\varphi\rangle\langle\varphi|\right)$. $\hat{\lambda }_{R, N}= \left | RN\right \rangle \left \langle RN\right |$, $R\in \left \{ H_{A_{1}},~H_{A_{2}},~V_{A_{4}},~V_{A_{2}}\right \}$, and $N\in \left \{ H_{B_{1}},~H_{B_{2}},~V_{B_{1}},~V_{B_{1}}\right \}$. Furthermore, one can obtain $\left|P_{i}\right|^{2},~\left|P_{2}\right|^{2},~\left|P_{3}\right|^{2},~\left|P_{4}\right|^{2},~\left|P_{5}\right|^{2},~\left|P_{6}\right|^{2},~\left|P_{7}\right|^{2}$, and $\left|P_{8}\right|^{2}$, i.e. $\left | P_{1}\right | ^{2}= \left \langle \hat{\lambda } _{H_{A_1}H_{B_1}}\right \rangle$, $\left | P_{2}\right | ^{2}= \left \langle \hat{\lambda }_{H_{A_1}V_{B_1}}\right \rangle$, $\left | P_{3}\right | ^{2}= \left \langle \hat{\lambda } _{V_{A_1}H_{B_1}}\right \rangle$, $\left | P_{4}\right | ^{2}= \left \langle \hat{\lambda }_{V_{A_1}V_{B_1}}\right \rangle$, $\left | P_{5}\right | ^{2}= \left \langle \hat{\lambda } _{H_{A_2}H_{B_2}}\right \rangle$, $\left |P_{6}\right | ^{2}= \left\langle\hat{\lambda} _{N_{A_2}V_{B_2}}\right \rangle$, $\left|P_{7}\right|^{2}=\left\langle\hat{\lambda} _{V_{A_2}H_{B_{2}}}\right \rangle$ and $\left|P_{8}\right|^{2}=\left\langle \hat{\lambda}_{V_{A_2}V_{B_{2}}}\right\rangle$. Thus, the probability amplitude of the 3-qubit quantum state, i.e., the linear measured signal $s_k=\left\{s_k^i\right\}_{i=1}^{d\times u}$ can be obtained. A specific description of the photon states and their projection basis is given in Appendix \ref{G}.

Finally, we discuss the preparation and the measurement method of $\left|Q_{k,nonlin}^{S}\right\rangle$. Despite that $\left|Q_{k,nonlin}^{S}\right\rangle$ is the tensor product of S states $\left|Q_{k,lin}\right\rangle$, to prepare the copies of $\left|Q_{k,lin}\right\rangle$ in experiments generally requires entangled photons, because no linear operators can copy an arbitrary $\left|Q_{k,lin}\right\rangle$ to another known state. Therefore, a quite many entangled photon pairs are required, which is consuming. Here, we consider a simplified version of achieving so, by repeating the preparation of $\left|Q_{k,lin}\right\rangle$ for $S$ times. Thus, by properly performing the required measurements on the states, $N_{_{r}}$ nonlinear measurement signals can be obtained. Particularly the results of the projection measurement can be denoted by $\left\langle\hat{\lambda}_{S}\right\rangle=tr\left(\hat{\lambda}_{S}\left|Q_{k,nonlin}^{S}\right\rangle\left|Q_{k,nonlin}^{S}\right\rangle\right)$, $\hat{\lambda }_{s}= | O\rangle \langle O|$, and
\begin{eqnarray}
    |O\rangle \in \{ \left|H_{A_{1}}H_{B_{1}}\right\rangle, \left | H_{A_{1}}V_{B_{1}}\right \rangle, \left|V_{A_{1}}H_{B_{1}}\right \rangle , \left|V_{A_{1}}V_{B_{1}}\right\rangle,\nonumber \\
    \left|H_{A_{2}}H_{B_{2}}\right \rangle, \left|H_{A_{2}}V_{B_{2}}\right \rangle, \left|V_{A_{2}}H_{B_{2}}\right\rangle, \left|V_{A_{2}}V_{B_{2}}\right \rangle \}^{\otimes S}.\nonumber
\end{eqnarray}
Finally, as described in the theoretical section, we perform a linear transformation on the linear and nonlinear measurement signals and obtain the corresponding weight matrix $W_{out}^{EX}$ under the experimental data by minimizing the mean square error.

Next, we show the results of the timer task and the Lorenzo63 task, respectively. In the timer task, we first encode l0 input signals onto a 3-qubit quantum state $\left|\varphi\right\rangle$, and prepare these l0 quantum states by adjusting the QHQs for $\hat{U}_{A},~\hat{U}_{B},~\hat{U}'_{A}$, and $\hat{U}'_{B}$ together with the HWPs in front of BBO. After that, we make projection measurements on the quantum states to obtain the probability amplitude of each of the 3-qubit quantum states. Next, we process the experimental data. In the training phase, we calculate the weight vector $W_{out}^{EX}$ of the linear transformation according to Eq.~(\ref{Eq.3}) and Eq.~(\ref{Eq.4}) Then, the processing results by $W_{out}^{EX}$ and by theoretically obtained $W_{out}$ are compared in the testing phase, displayed in Fig.~\ref{Fig.6}(a). The orange solid line indicates the experimentally tested results with time steps, and the blue solid lines indicates the theoretically tested results. We can see that at time step 9, both results have a timed-out response, indicating that the weight vector $W_{out}^{EX}$ obtained experimentally can be effectively applied to the timer task.
\begin{figure}
    \centering
    \includegraphics{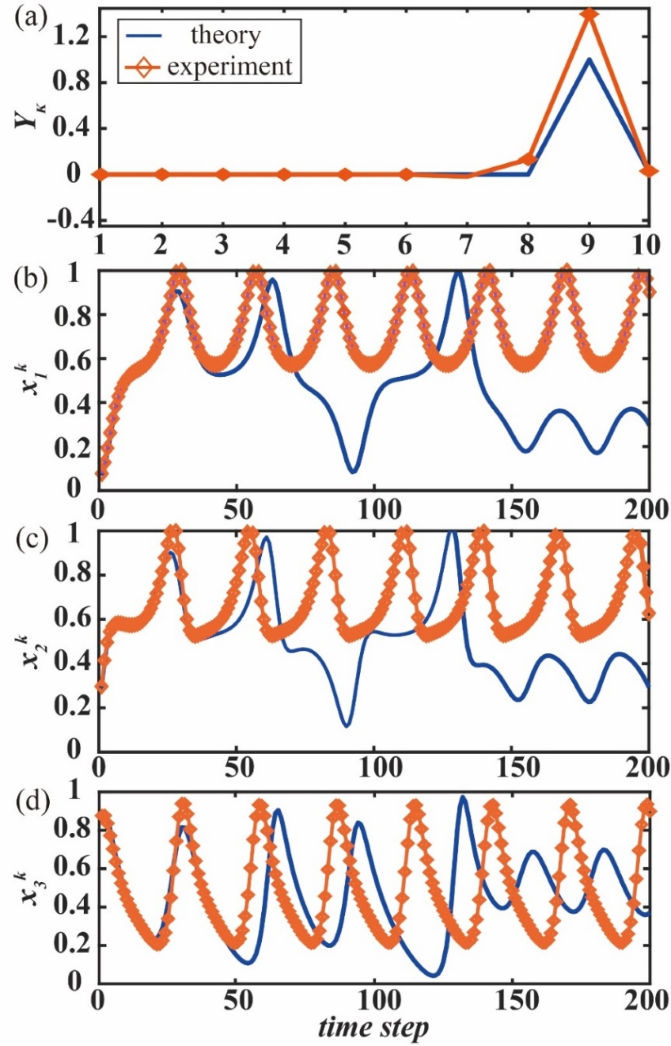}
    \caption{The experimental performance of QRC with nonlinear vector autoregression in the timer task (a) and the Lorenz63 prediction task (b), (c) and (d). The experiment curve during the testing phase is colored orange. The numerical results of theory are colored blue, serving as a reference.}
    \label{Fig.6}
\end{figure}

In the Lorenzo63 task, we process 400 quantum states of 3 qubits. First, we still encode the 400 input signals by the 3-qubit quantum state $|\varphi\rangle$, and prepare all quantum states by adjusting the waveplates on the optical path. Next, we do projection measurements on copies of $|\varphi\rangle$ to obtain the amplitudes. In the task, one copy of the state is needed, so we prepare the state twice to obtain the measurement results. After that, we process the experimental data. In the training phase, we also calculate the weight matrix $W_{out}^{EX}$ and test the numerical results by the $W_{out}^{EX}$ and the $W_{out}$ in the testing phase, and the results of the two parts are shown in Fig.~\ref{Fig.6}(b), (c), and (d), respectively. The orange solid lines show the experimental test results $x_1^k$, $x_2^k$ and $x_3^k$ as a function of time step, while the blue solid lines show the theoretical test results. We can see that the experimental results of $x_1^k$, $x_2^k$, and $x_3^k$ are in good agreement with the theoretical results at the beginning, indicating that the weight matrix $W_{out}^{EX}$ obtained from the experiment can be effectively applied to the Lorenzo63 task. However, when $k=50$, the experimental results begin to show significant errors with the theoretical results. This means the prediction time step is shorter compared to the numerical results shown in Fig.~\ref{Fig.3}. We believe that it is caused by the imperfections of the experimental system.

Last but not least, we analysis the noise in our experiment more generally. It is shown that the physical implementation of QRC is sensitive to system noise, and the affection of noise is complicated \cite{Domingo_2023}. Based on the setup of our scheme, the coincidence count per second can reach a relatively high level. Therefore, our experimental device is not disturbed much by background noise. The major noise of our experiments comes from the generation of entangled state. \red{In fact, the generation of entangled state is not perfect, and we evaluate the quality of the generated state by the visibility $V=(C_{\max}-C_{\min})/(C_{\max}+C_{\min})$, where $C_{\max}$ is the maximum coincidence counts and $C_{\min}$ is the minimum coincidence counts. (see more details in Appendix ~\ref{E} ).} Compared to the fidelity based on state tomography, this quantity is easier to measure and calculate, and effective to explain the results. If the visibility is close to 1, the prepared entangled state will be close to the target. In the experiments, because the encoding of input data is realized by applying unitary operators to the entangled states, the visibility of the states fundamentally determines the quality of data processing. Specifically, for the Bell state $(|HH\rangle+|VV\rangle)/\sqrt{2}$, if the visibility of the prepared state is not 100$\%$, it indicates that the state components such as $|HV\rangle$ or $|VH\rangle$ are also generated. This leads to a global error, which transmits to all the elements of data points by those unitary operators. It is known that the dynamics of the Lorenzo63 system is chaotic, and is very sensitive to the fluctuation of parameters. Therefore, the global error, induced by the imperfection of the generated entangled state, results in an unpredictable deviation from the numerical calculation. Such a deviation is similar to the fluctuation of parameters, and further affect the advance in data efficiency for training. Our results also indicates that the above affection cannot be eliminated by linear fitting. A potentially better implementation could be optical chip or other integrated optics designs, as they have controls with a higher quality than the bulk optics.

\section{\label{sec:5}CONCLUSION }

We have proposed a theoretical QRC scheme with nonlinear vector autoregression. Our QRC scheme discards the traditional framework of the reservoir as a neural network, adopts merely quantum states as the reservoir that gives the linear and nonlinear features of data. We have effectively performed numerical simulations for the timer task, Lorenzo63 prediction task, and several other tasks. Compared with the traditional QRC scheme, our proposal does not involve an injection of the data during the evolution of a quantum system, which leads to both the suppression of costs in warm-up stage and the ease of experimental implementation. Furthermore, an improvement in the amount of training data for obtaining a reliable outcome can be observed, which benefits the current study of complex dynamical systems. Finally, we have realized the preparation of arbitrary 3-qubit quantum states on the quantum optics platform, presenting a new idea for the experimental realization of the QRC scheme. In the future exploration, our scheme is expected to be helpful for investigating other complex chaotic system models. It would a more efficient strategy for performing numerical simulations and experimental demonstrations of the systems. 

\begin{acknowledgments}
This work was supported by the National key R \& D Program of China under Grant No.2022YFA1404904 and the National Natural Science Foundation of China (No. 12234004 and No.11904022).
\end{acknowledgments}

\appendix

\section{\label{A}THE NARMA NONLINEAR DYNAMICAL SYSTEM}

This task is to simulate nonlinear dynamical systems known as NARMA, which is a standard benchmark task in the context of recurrent neural network learning. Due to its nonlinearity and dependence on long time lags, this task is a challenging problem for any computing system. We use a superimposed sine wave as input to the NARMA system, and are written as 
\begin{equation}
    x_k=0.1\left[\sin(\frac{2\pi\alpha k}{T})\sin(\frac{2\pi\beta k}{T})\sin(\frac{2\pi\gamma k}{T})+1\right],
\end{equation}
where $(\alpha,~\beta,~\gamma)=(2.11,~3.73,~4.11)$ and $T=100$. The first NARMA system, NARMA2, is the following second-order nonlinear dynamical system
\begin{equation}
    y_{k+1}=0.4y_{k}+0.4y_{k}y_{k-1}+0.6\left(x_{k}\right)^{3}+0.1.
\end{equation}

The second NARMA system is the following nonlinear dynamical system, which is of order n,
\begin{equation}
    y_{k+1}=\alpha y_{k}+\beta\Bigg(\sum_{j=0}^{n-1}y_{k-j}\Bigg)+\gamma x_{k-n+1}x_{k}+\delta .
\end{equation}
$(\alpha,~\beta,~\gamma,~\delta)=(0.3,~0.05,~1.5,~0.1)$. Here, we consider four cases when values of $n$ are 5, 10, 15, and 20. The corresponding systems are termed by NARMA5, NARMA10, NARMA15, and NARMA20, respectively. In fact, the investigation on NARMA systems do not need warm-up data. As shown in Table S1, we show the setup of quantum states for each NARMA system separately. Further, we numerically simulated each NARMA task separately, and the results are shown in Fig. \ref{Fig.7}(a) to (e). Each of the panel in Fig. \ref{Fig.7} shows the variation of the output variable   of each task over time steps in the testing phase. The target signals are colored blue and the predicted signal are colored orange, respectively. We can clearly see that the consistency between the predicted signal and the target signal is very good. Further, we calculated their normalized root-mean-square (NRMSE) error separately. They are $5.6\times10^{-4}$, $3.0\times10^{-3}$, $2.1\times10^{-3}$, $5.9\times10^{-3}$ and $9.6\times10^{-3}$. Performing the NARMA task in the traditional QRC scheme \cite{Fujii_2017} requires 4000 datasets to train the model. In this scheme only 1000 or 2000 datasets are required to train the model. Compared to the traditional QRC scheme, this scheme can efficiently predict the NARMA task with a small number of training datasets and a two to four times reduction in the training datasets, while realizing a smaller error.
\begin{table}[htbp]
\caption{\label{table1}
The parameters and NRMSE errors of each NARMA task.}
\begin{ruledtabular}
\begin{tabular}{ccccccc}
 Task name&Training size &Test size&
 $q$  &$u$ & $S$ & Error\\
\hline
NARMA2& 1000 & 200 & 1 &8
& 4 & $5.6\times10^{(-4)}$ \\
NARMA5& 1000 & 200 & 1 &8
& 4 & $3.0\times10^{(-3)}$ \\
NARMA10& 1000 & 200 & 1 &8
& 4 & $2.1\times10^{(-3)}$ \\
NARMA15& 1000 & 200 & 1 &8
& 4 & $5.9\times10^{(-3)}$ \\
NARMA20& 2000 & 200 & 16 &8
& 4 & $9.6\times10^{(-3)}$ \\
\end{tabular}
\end{ruledtabular}
\end{table}

\begin{figure}
    \centering
    \includegraphics{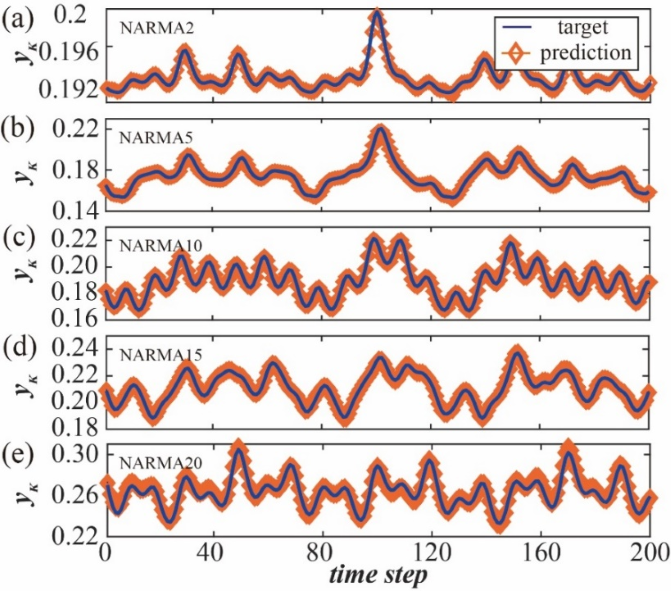}
    \caption{Prediction results of the NARMA task. (a) to (e) shows the variation of the output variable  $y_k$ of each task over time steps in the testing phase, and the specific task names are marked in the corresponding figure. The target signals are colored blue and the predicted signal are colored orange, respectively. }
    \label{Fig.7}
\end{figure}

\section{\label{B}THE MACKEY-GLASS (MG) TIME SERIES PREDICTION TASK}
This task is the MG time series prediction task, which is a chaotic time series. This is also a well-known benchmark task in machine learning. The MG prediction task is determined by a lagged differential equation, which can be represented as
\begin{equation}
    \dot{y}\left(t\right)=\frac{\alpha y\left(t-\tau_{MG}\right)}{1+y\left(t-\tau_{MG}\right)^\beta}-\gamma y\left(t\right),
\end{equation}
where $\left(\alpha,~\beta,~\gamma\right)=\left(0.2,~10,~0.1\right )$. When $\tau_{MG}>16.8$, the system has a chaotic attractor. In most studies, one chooses $\tau_{MG}=17$ to study the chaos of the system \cite{Spagnolo_2022}. Similarly, this task uses the parameter setting $\tau_{MG}=17$. For the MG prediction task, we generate 1700 normalized input data samples through numerical integration. The first 200 data samples are used for flushing, followed by 1000 data samples for the training phase and 500 data samples for the testing phase. The states for giving the features are set as follows. we set the state $|Q_{k,lin}\rangle=\begin{pmatrix}x_k,~x_{k-1},~x_{k-2},~x_{k-3},~x_{k-4},~x_{k-5},~x_{k-6},~x_{k-7}\end{pmatrix}/\sqrt{N}$ for giving the linear features. We also set $S=5$ for state $\left|Q^S_{k,nonlin}\right\rangle$, i.e., $\left|Q_{k,nonlin}^5\right\rangle=\left|Q_{k,lin}\right\rangle\otimes\left|Q_{k,lin}\right\rangle\otimes\left|Q_{k,lin}\right\rangle\otimes\left|Q_{k,lin}\right\rangle\otimes\left|Q_{k,lin}\right\rangle$ Next, the MG prediction task is still accomplished by numerical simulation. Due to the sensitivity of the chaotic model to data, after obtaining the measurement results, this scheme multiplies the normalization coefficient $\sqrt{N}$ of the quantum state with the measurement results. The results are shown in Fig.~\ref{Fig.8}, which also illustrates the variation of $y_k$ over time steps in the test phase. We found that the target data and the predicted data at time steps 1 to 300 match very well in the test phase. Further, we calculated the NRMSE of the data when the time step takes from 1 to 300, and the result is 0.16. Performing the MG prediction task in the traditional QRC scheme \cite{Fujii_2017} requires 10,000 datasets to train the model. While in the above scheme only 1200 datasets are required to train the model. Compared to traditional QRC scheme, this method can effectively predict complex chaotic systems with a small amount of training data, even with relatively small errors, and reduce the training dataset by an order of magnitude.

\begin{figure}
    \centering
    \includegraphics{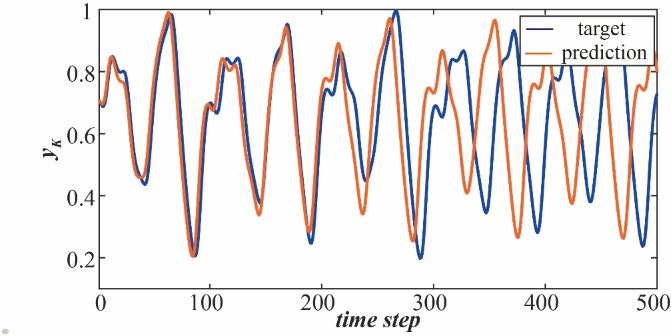}
    \caption{Prediction results of the MG task. The target signals are colored blue and the predicted signal are colored orange, respectively.}
    \label{Fig.8}
\end{figure}
\section{\label{C}A DOUBLE-SCROLL ELECTRONIC CIRCTUI (DSEC) PREDICTION TASK}
This task is the DSEC prediction task. The DSEC model serves as a typical benchmark task demonstrating the chaotic behavior of a circuit system. DSEC is composed of three coupled nonlinear differential equations, which can be represented as 
\begin{eqnarray}
    &&\dot{x}_{1}=x_{1}/D_{1}-\Delta x_{12}/D_{2}-2D_{5}\sinh(D_{4}\Delta x_{12}), \nonumber\\
&&\dot{x}_{2}=\Delta x_{12}/D_{2}+2D_{5}\sinh(D_{4}\Delta x_{12})-x_{3}, \nonumber\\
&&\dot{x}_{3}=x_{2}-D_{3}x_{3},
\end{eqnarray}
where $D_1=1.2,~D_2=3.44,~D_3=0.193,~D_4=11.6,~D_5=2.25\times10^{-5},~\mathrm{and}~\Delta V=V_1-V_2$. 
For the DSEC prediction task, we generated 4500 normalized input data samples through numerical integration. The first 200 data samples are used for flushing, followed by 3800 data samples for the training phase and 500 data samples for the testing phase. Next, according to the above theoretical scheme, the parameters of the quantum state for linear feature are set by $q=1$ and $u=8$, and the parameters of the quantum state for nonlinear feature are set by $S=3$. As such, the size of the state for linear feature is a 3-qubit state, while the one for nonlinear feature is a 9-qubit state. Therefore, as we have demonstrated, a 9-qubit quantum state is sufficient to build a QR as effective as a nonlinear vector autoregression with 62 features \cite{Gauthier_2021}. Similarly, we perform the DSEC prediction task by numerical simulation. We multiply the normalization coefficient $\sqrt{N}$ of the quantum state with the measurement results and then perform the linear transformation operation. The numerical results are shown in Fig.~\ref{Fig.9}, which illustrates the variation of $x_1^k$, $x_2^k$, and $x_3^k$ with time steps during the test phase, respectively. We computed the NRMSE for the target and predicted data for time steps when the time step takes from 1 to 135 in the testing phase. They are $2.1\times 10^{-2}$, $3.2\times 10^{-2}$ and $2.2\times 10^{-2}$ The scheme also achieves the effective prediction of complex chaotic systems with a small amount of training dataset under an acceptable error bound.
\begin{figure}
    \centering
    \includegraphics{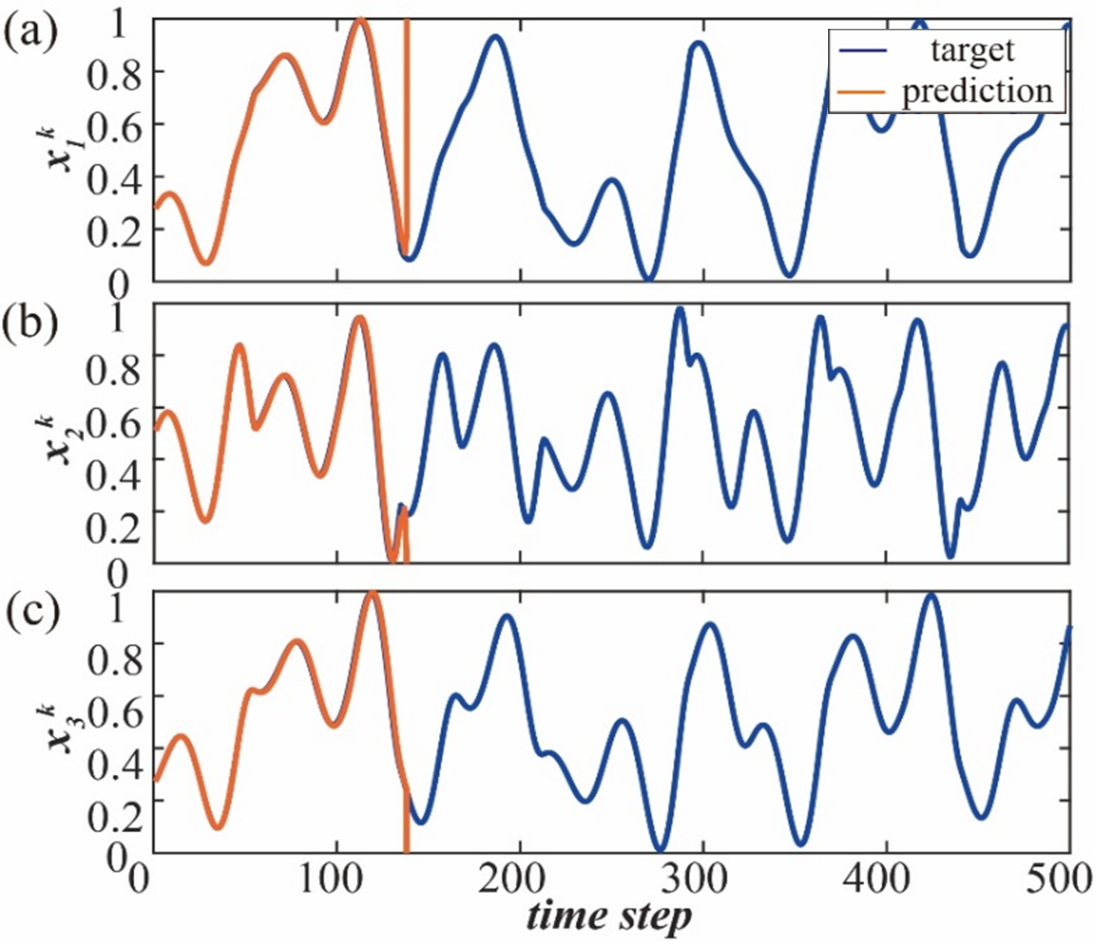}
    \caption{Prediction results of the DSEC prediction task. The target signals are colored blue and the predicted signal are colored orange, respectively. }
    \label{Fig.9}
\end{figure}
\section{\label{D}THE TUNING OF RATIO BETWEEN $\alpha'$ AND $\beta'$ BY THE HWP BEFORE THE PBS}
In the experimental setup shown by Fig.~\ref{Fig.5}, we use a weak focusing method to pump the Type-I BBO, and then modulate the power of the pump light to achieve modulation of the final coincidence counts. First, we pump the BBO without the measurement basis to directly collect the entangled photons and record the coincidence counts. We find that the power of pump light varies from 0-700mW, which is linearly related to the coincidence counts of the photons pumped by BBO. This means that if one tunes the HWP before the PBS, the power of each beam output by PBS can be correspondingly tunned. Hence, the ratio of the counts of the entangled photon pairs output by two BBOs, which corresponds to the ratio of $\alpha'$ and $\beta'$ in Eq.~(\ref{Eq.9}), can be effectively tuned. As shown in Fig.~\ref{Fig.10} below, according to the fitting results, the coincidence counts of the down conversion of the two BBOs are approximately given
\begin{eqnarray}
    &C_{1}=11.43*P_{1},\\
    &C_{2}=11.91*P_{2},
\end{eqnarray}
where $P_1$ and $P_2$ are the pump power of the two BBOs respectively. The unit of the two coefficients are $W^{-1}$.
\begin{figure}
    \centering
    \includegraphics{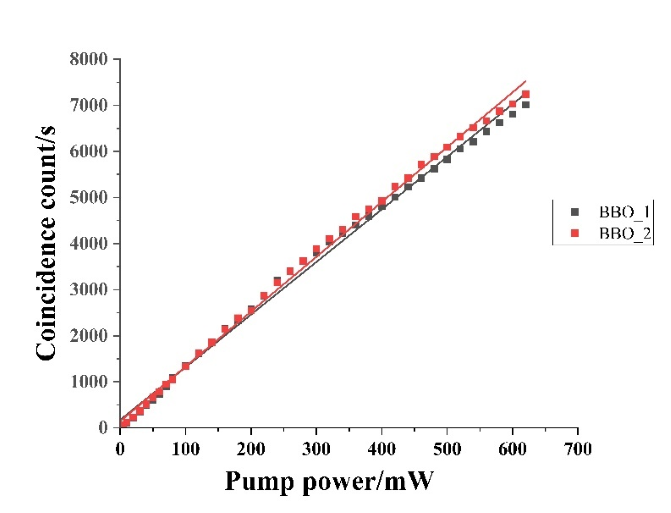}
    \caption{Schematic representation of the linear relationship between power and coincidence counts for Pump light. The red solid line indicates the schematic diagram of BBO’s coincidence counts in light path I as a function of power. The black solid line indicates the schematic diagram of the BBO's coincidence counts in light path II as a function of power.}
    \label{Fig.10}
\end{figure}

\section{\label{E}THE MEASUREMENT OF POLARIZATION CORRELATION CURVES}

In order to ensure that the entangled sources pumped by the two BBOs are effective to be used under the experimental conditions, we separately hit the two BBOs with pump light under random conditions and measure their polarization correlation curves respectively. The results are shown in Fig.~\ref{Fig.11} below. It shows the coincidence counts as a function of the signal polarization analyzer angle $\theta_1$ when the idler polarization analyzer angle $\theta_2$ was fixed at two different values $0^{\circ}$ and $45^{\circ}$. We used $\hat{U}_{A},~\hat{U}_{B},~\hat{U}_{A}^{\prime},~\hat{U}_{B}^{\prime}$ and HWP in front of BBOs to set the entanglement state as $\left|\psi\right\rangle=\frac{1}{\sqrt{2}}\left(\left|HH\right\rangle+\left|VV\right\rangle\right)$. For input pump power of 309 mW in path I and 425 mW in path II, we measured two groups of polarization correlation curves respectively. quantum-interference fringe visibility $V$ is an important parameter to measure the clarity of interference patterns in quantum interference experiments. It can be defined by maximum coincidence counts ($C_{\max}$) and minimum coincidence counts ($C_{\min}$) at a certain measuring base. Specifically, it can be calculated by the following formula 
\begin{equation}
    \red{V=\frac{(C_{\max}-C_{\min})}{(C_{\max}+C_{\min})}.}
\end{equation}
\begin{figure}[htbp]
    \centering
    \includegraphics[width=3.4in]{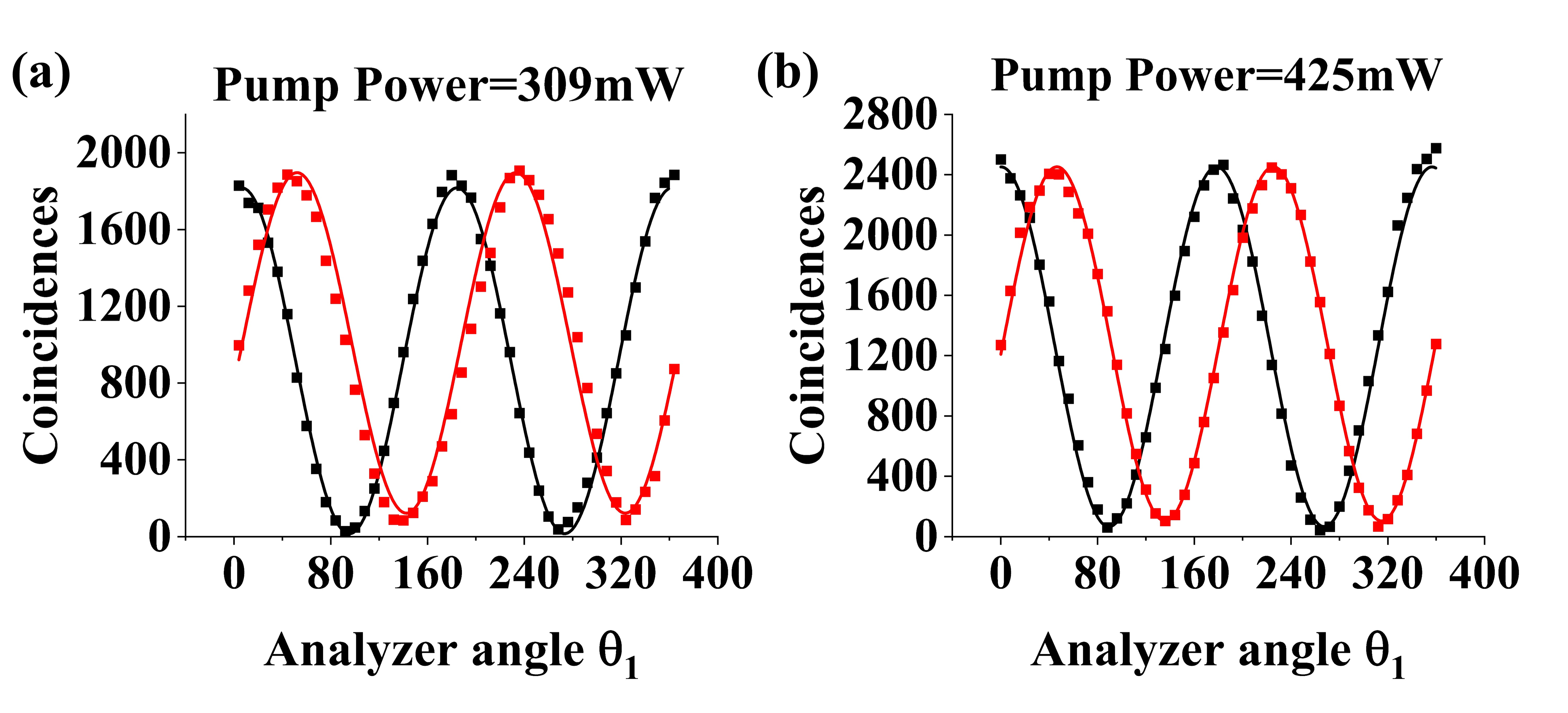}
    \caption{Coincidence counts as a function of signal polarization analyzer angle $\theta_1$ for different settings of idler polarization analyzer angle $\theta_2$: $0^{\circ}$ (black squares), $45^{\circ}$ (red squares). Solid lines are best sinusoidal fits to data.}
    \label{Fig.11}
\end{figure}

We calculate the visibilities for the two sets of measurement in path I, which is $V_{0^{\circ}}^{\text{pathI}}=96.98\pm0.6$, and $V_{45^{\circ}}^{\text{pathI}}=94.62\pm1.55$. The visibilities for the two sets of measurement in path II are $V_{0^{\circ}}^{\text{pathII}}=97.47\pm1.2$ and $V_{45^{\circ}}^{\text{pathII}}=94.62\pm0.71$.

\section{\label{F}REALIZATION OF THE FOUR ROTATION MATRICES $\hat{U}_{A},~\hat{U}_{B},~\hat{U}_{A}',~\hat{U}_{B}'$}
Now we discuss how to use wave plates to realize four rotation matrices $\hat{U}_{A},~\hat{U}_{B},~\hat{U}_{A}',~\hat{U}_{B}'$. According to Ref.~\cite{Kwiat_1999} in the main text, the two-photon state
\begin{equation}
a|HH\rangle+b|HV\rangle+c|VH\rangle+d|VV\rangle,
\end{equation}
can be implemented by applying $\hat{U}_A\otimes\hat{U}_B$ to $\alpha|HH\rangle+\beta|VV\rangle$. $\hat{U}_A~\mathrm{and}~\hat{U}_B$ are defined by
\begin{eqnarray}
    \hat{U}_{A}=\begin{pmatrix}u_1&&v_1\\-v_1^*&&u_1^*\end{pmatrix},\label{F2}\\\hat{U}_{B}=\begin{pmatrix}u_2&&v_2\\-v_2^*&&u_2^*\end{pmatrix}.\label{F3}
\end{eqnarray}
$\alpha$ and $\beta$ satisfy that
\begin{eqnarray}
    &&\alpha=\frac{\sqrt{1-\sqrt{1-4\left|ad-bc\right|^{2}}}}{\sqrt{2}} ,\\&&\beta=\frac{ad-bc}{a},
\end{eqnarray}
where we restrict $ad-bc\neq0$ and $\left|ad-bc\right|\neq\frac12$. The parameters in Eq.~(\ref{F2}) and Eq.~(\ref{F3}) are given by 
\begin{eqnarray}
    u_1\equiv\frac{\left|z_1\right|}{\sqrt{\left|z_1\right|^2-\left|z_3\right|^2}},v_1\equiv\frac{z_3u_1^*}{\left|z_1^*\right|},\nonumber\\
    u_2\equiv\frac{z_3^*}{z_1z_3+z_2z_4},v_2\equiv\frac{z_2}{z_1z_3+z_2z_4},
\end{eqnarray}
where $z_{1}=\frac{a\alpha^{*}-d^{*}\beta}{\left|\alpha\right|^{2}-\left|\beta\right|^{2}},z_{2}=\frac{d^{*}\alpha-\beta^{*}\alpha}{\left|\alpha\right|^{2}-\left|\beta\right|^{2}},z_{3}=\frac{-\alpha c^{*}-\beta^{*}b}{\left|\alpha\right|^{2}-\left|\beta\right|^{2}},z_{4}=\frac{-b\alpha^{*}-c^{*}\beta}{\left|\alpha\right|^{2}-\left|\beta\right|^{2}}$. 
On the one hand, the matrix form of $\hat{U}_A$ and $\hat{U}_B$ corresponds exactly to the Jones matrix of the half-wave piece (HWP), which is
\begin{equation}
    U_{HWP}\big(\theta\big)=-i\bigg(\begin{matrix}\cos(2\theta)&\sin(2\theta)\\\sin(2\theta)&-\cos(2\theta)\end{matrix}\bigg).
\end{equation}
Therefore, $\hat{U}_A$ and $\hat{U}_B$ can be implemented by HWPs. There are a few differences in the factors of the matrix elements, but they have no affections on the projection measurement in our consideration. Therefore, we can set that
\begin{equation}
    \begin{split}
    \hat{U}_A=&\begin{pmatrix}u_1&v_1\\-v_1^*&u_1^*\end{pmatrix}=U_{HWP}\begin{pmatrix}\theta\end{pmatrix}\\
    =&
    -i\begin{pmatrix}\cos(2\theta)&\sin(2\theta)\\\sin(2\theta)&-\cos(2\theta)\end{pmatrix},    
    \end{split}
\end{equation}
i.e., the angle of the HWP is $\theta_{\hat{U}_A}=\frac12\arccos u_1$, for a given $\hat{U}_A$. Similarly for $\hat{U}_B$, one has $~\theta_{\hat{U}_B}=\frac12\arccos u_2$.

On the other hand, $\alpha$ and $\beta$ can be controlled by the HWP in front of the BBO, i.e., pump light whose polarization state is $\alpha|V\rangle+e^{i\phi}\beta|H\rangle$ produces photon pairs with $\alpha|HH\rangle+e^{i\phi}\beta|VV\rangle$ \cite{devaney2018introduction}. Then, we can calculate the angle $\theta$ of the HWP in front of the BBO according to the Jones matrix. $\theta=\frac{1}{2}\arctan\left(\frac{\beta}{\alpha}\right)$ when the Pump light is $|V\rangle$, $\theta=\frac{1}{2}\arctan\left(\frac{\alpha}{\beta}\right)$ when the Pump light is $|H\rangle$.

\section{\label{G}THE PHOTON STATES AND THEIR PROJECTION BASIS IN OUR EXPERIMENTS}
After obtaining the entangled photon pairs, the two pairs of photon states after being manipulated by $\hat{U}_{A},~\hat{U}_{B},~\hat{U}_{A}',~\hat{U}_{B}'$ can be given by
\begin{eqnarray}
\hat{U}_{A}\otimes\hat{U}_{B}&&\left(\cos\theta_{1}\left|H_{A_1}H_{B_1}\right|+e^{i\phi_1}\sin\theta_{_1}\left|V_{_{A_1}}V_{_{B_1}}\right|\right)\nonumber \\
=&&a\left|H_{A_{1}}H_{B_{1}}\right\rangle+b\left|H_{A_{1}}V_{B_{1}}\right\rangle+\nonumber\\ &&c\left|V_{A_{1}}H_{B_{1}}\right\rangle+d\left|V_{A_{1}}V_{B_{1}}\right\rangle,
\end{eqnarray}

\begin{eqnarray}
\hat{U}_{A}^{\prime}\otimes\hat{U}_{B}^{\prime}&&\left(\cos\theta_{2}\left|H_{A_{2}}H_{B_{2}}\right|+e^{i\phi_{2}}\sin\theta_{2}\left|V_{A_{2}}V_{B_{2}}\right\rangle\right) \nonumber\\
=&&a^{\prime}\left|H_{A_{2}}H_{B_{2}}\right\rangle+b^{\prime}\left|H_{A_{2}}V_{B_{2}}\right\rangle+\nonumber\\
&&c^{\prime}\left|V_{A_{2}}H_{B_{2}}\right\rangle+d^{\prime}\left|V_{A_{2}}V_{B_{2}}\right\rangle.
\end{eqnarray}
Further, state $|\varphi\rangle$ (Eq.~(\ref{Eq.9})) can be given by using the above two pairs of photon states. The projective measurement operation on the state $|\varphi\rangle$ can be given by 
\begin{equation}
    \begin{cases}
    \left\langle\hat{\lambda}_{H_{A_{1}}H_{B_{1}}}\right\rangle=tr\left(\hat{\lambda}_{H_{A_{1}}H_{B_{1}}}\left|Q_{k,lin}\right\rangle\langle Q_{k,lin}|\right)\\
    \left\langle\hat{\lambda}_{H_{A_{1}}V_{B_{1}}}\right\rangle=tr\left(\hat{\lambda}_{H_{A_{1}}V_{B_{1}}}\left|Q_{k,lin}\right\rangle\langle Q_{k,lin}|\right)\\
    \left\langle\hat{\lambda}_{V_{A_{1}}H_{B_{1}}}\right\rangle=tr\left(\hat{\lambda}_{V_{A_{1}}H_{B_{1}}}\left|Q_{k,lin}\right\rangle\langle Q_{k,lin}|\right)\\
    \left\langle\hat{\lambda}_{V_{A_{1}}V_{B_{1}}}\right\rangle=tr\left(\hat{\lambda}_{V_{A_{1}}V_{B_{1}}}\left|Q_{k,lin}\right\rangle\langle Q_{k,lin}|\right)\\
    \left\langle\hat{\lambda}_{H_{A_{2}}H_{B_{2}}}\right\rangle=tr\left(\hat{\lambda}_{H_{A_{2}}H_{B_{2}}}\left|Q_{k,lin}\right\rangle\langle Q_{k,lin}|\right)\\
    \left\langle\hat{\lambda}_{H_{A_{2}}V_{B_{2}}}\right\rangle=tr\left(\hat{\lambda}_{H_{A_{2}}V_{B_{2}}}\left|Q_{k,lin}\right\rangle\langle Q_{k,lin}|\right)\\
    \left\langle\hat{\lambda}_{V_{A_{2}}H_{B_{2}}}\right\rangle=tr\left(\hat{\lambda}_{V_{A_{2}}H_{B_{2}}}\left|Q_{k,lin}\right\rangle\langle Q_{k,lin}|\right)\\
    \left\langle\hat{\lambda}_{V_{A_{2}}V_{B_{2}}}\right\rangle=tr\left(\hat{\lambda}_{H_{V_{2}}V_{B_{2}}}\left|Q_{k,lin}\right\rangle\langle Q_{k,lin}|\right).\\
    \end{cases}
\end{equation}

As we described in the main text, the amplitudes of $|\varphi\rangle$ can be obtained by the above measurements. It is worth mentioning that, in our experimental setup, the measurement basis is relatively simple. Therefore, the measurements on each pair of photons can be performed independently. For the case involving more complex basis state, such as the superposition of the basis states in our scheme, the projection on them can be done by firstly projecting the state on simple basis and secondly adding up the projection results accordingly to give the final outcome.


\begin{thebibliography}{99}
\bibitem{Jordan_2015}
M. I. Jordan and T. M. Mitchell, Machine learning: trends, perspectives, and prospects, \sci{349}, 255-260 (2015).
  
\bibitem{bishop2006pattern}
C. M. Bishop and N. M. Nasrabadi, \emph{Pattern recognition and machine learning} (Springer New York, NY, 2006)

\bibitem{Jaeger_2004}
H. Jaeger and H. Haas, Harnessing nonlinearity: predicting chaotic systems and saving energy in wireless communication, \sci{304}, 78-80 (2004).

\bibitem{Milano_2021}
G. Milano, G. Pedretti, K. Montano, S. Ricci, S. Hashemkhani, L. Boarino, D. Ielmini, and C. Ricciardi, In materia reservoir computing with a fully memristive architecture based on self-organizing nanowire networks, Nat. Mater. \textbf{21}, 195–202 (2021). 

\bibitem{Pathak_2018}
J. Pathak, B. Hunt, M. Girvan, Z. Lu, and E. Ott, Model-Free Prediction of large spatiotemporally chaotic systems from data: a reservoir computing approach, \prl{120}, 024102 (2018).

\bibitem{Gauthier_2021}
D. J. Gauthier, E. Bollt, A. Griffith, and W. A. S. Barbosa, Next generation reservoir computing, Nat. Commun. \textbf{12}, 5564 (2021). 

\bibitem{Tanaka_2019}
G. Tanaka, T. Yamane, J. B. Héroux, R. Nakane, N. Kanazawa, S.Takeda, H. Numata, D. Nakano, and A. Hirose, Recent advances in physical reservoir computing: A review, Neural Netw. \textbf{115}, 100-123 (2019).

\bibitem{Yan_2022}
X. Yan, Reservoir computing goes fully analogue, Nat. Electron. \textbf{5}, 629-630 (2022).

\bibitem{Cucchi_2021}
M. Cucchi, C. Gruener, L. Petrauskas, P. Steiner, H. Tseng, A. Fischer, B. Penkovsky, C. Matthus, P. Birkholz, H. Kleemann, and k. Leo, Reservoir computing with biocompatible organic electrochemical networks for brain-inspired biosignal classification, Sci. Adv. \textbf{7}, eabh0693 (2021).

\bibitem{Zhong_2022}
Y. Zhong, J. Tang, X. Li, X. Liang, Z. Liu, Y. Li, Y. Xi, P. Yao, Z. Hao, B. Gao, H. Qian, and H. Wu, A memristor-based analogue reservoir computing system for real-time and power-efficient signal processing, Nat. Electron. \textbf{5}, 672-681 (2022).

\bibitem{Chen_2023}
Z. Chen, W. Li, Z. Fan, S. Dong, Y. Chen, M. Qin, M. Zeng, X. Lu, G. Zhou, X. Gao, and J.-M. Liu, All-ferroelectric implementation of reservoir computing, Nat. Commun. \textbf{14}, 3585 (2023).

\bibitem{Tan_2023}
H. Tan and S. van Dijken, Dynamic machine vision with retinomorphic photomemristor-reservoir computing, Nat. Commun. \textbf{14}, 2169 (2023).

\bibitem{Zhong_2021}
Y. Zhong, J. Tang, X. Li, B. Gao, H. Qian, and H. Wu, Dynamic memristor-based reservoir computing for high-efficiency temporal signal processing, Nat. Commun. \textbf{12}, 408 (2021).

\bibitem{Vandoorne_2014}
K. Vandoorne, P. Mechet, T. Van Vaerenbergh, M. Fiers, G. Morthier, D. Verstraeten, B. Schrauwen, J. Dambre, and P. Bienstman, Experimental demonstration of reservoir computing on a silicon photonics chip, Nat. Commun. \textbf{5}, 3541 (2014).

\bibitem{Sun_2021}
L. Sun, Z. Wang, J. Jiang, Y. Kim, B. Joo, S. Zheng, S. Lee, W.-J. Yu, B.-S. Kong, and H. Yang, In-sensor reservoir computing for language learning via two-dimensional memristors, Sci. Adv. \textbf{7}, eabg1455 (2021).

\bibitem{Appeltant_2011}
L. Appeltant, M. C. Soriano, G. Van der Sande,  J. Danckaert, S. Massar, J. Dambre, B. Schrauwen, C. R. Mirasso, and I. Fischer, Information processing using a single dynamical node as complex system, Nat. Commun. \textbf{2}, 468 (2011).

\bibitem{Rafayelyan_2020}
M. Rafayelyan, J. Dong, Y. Tan, F. Krzakala, and S. Gigan, Large-scale optical reservoir computing for spatiotemporal chaotic systems prediction, Phys. Rev. X \textbf{10}, 041037 (2020).

\bibitem{Liu_2022}
K. Liu, B. Dang, T. Zhang, Z. Yang, L. Bao, L. Xu, C. Cheng, R. Huang, and Y. Yang, Multilayer reservoir computing based on ferroelectric $\alpha$‐$\rm{In}_2\rm{Se}_3$ for hierarchical information processing, Adv. Mater. \textbf{34}, 48, 2270333 (2022).

\bibitem{Brunner_2013}
D. Brunner, M. C. Soriano, C. R. Mirasso, and I. Fischer, Parallel photonic information processing at gigabyte per second data rates using transient states, Nat. Commun. \textbf{4}, 1364 (2013).

\bibitem{Woods_2012}
D. Woods and T. J. Naughton, Photonic neural network, Nat. Phys. \textbf{8}, 257–259 (2012).

\bibitem{Du_2017}
C. Du, F. Cai, M. A. Zidan, W. Ma, S. Lee, and W. D. Lu, Reservoir computing using dynamic memristors for temporal information processing, Nat. Commun. \textbf{8}, 2204 (2017).

\bibitem{Wei_2022}
Z. Wei, Reservoir computing with 2D materials, Nat. Electron. \textbf{5}, 715–716 (2022). 

\bibitem{Moon_2019}
J. Moon, W. Ma, J.-H. Shin, F. Cai, C. Du, S.-H. Lee, and W.-D. Lu, Temporal data classification and forecasting using a memristor-based reservoir computing system, Nat. Electron. \textbf{2}, 480–487 (2019).

\bibitem{Akashi_2022}
N. Akashi, Y. Kuniyoshi, S. Tsunegi, T. Taniguchi, M. Nishida, R. Sakurai, Y. Wakao, K. Kawashima, and K. Nakajima, Coupled spintronics neuromorphic approach for high‐performance reservoir computing, Adv. Intell. Syst. \textbf{4}(10), 2200123 (2022).

\bibitem{domingo2023anticipating}
L. Domingo, M. Grande, F. Borondo, and J. Borondo, Anticipating food price crises by reservoir computing, Chaos T. Fractal \textbf{174}, 113854 (2023).

\bibitem{Nakajima_2019}
K. Nakajima, K. Fujii, M. Negoro, K. Mitarai, and M. Kitagawa, Boosting computational power through spatial multiplexing in quantum reservoir computing,  Phys. Rev. Applied \textbf{11}, 034021 (2019).

\bibitem{Spagnolo_2022}
M. Spagnolo, J. Morris, S. Piacentini, M. Antesberger, F. Massa, A. Crespi, F. Ceccarelli, R. Osellame, and P. Walther, Experimental photonic quantum memristor, Nat. Photon. \textbf{16}, 318–323 (2022).

\bibitem{Fujii_2017}
K. Fujii and K. Nakajima, Harnessing disordered-ensemble quantum dynamics for machine learning, Phys. Rev. Applied \textbf{8}, 024030 (2017).

\bibitem{tran2020higher}
Q.-H. Tran and K. Nakajima, Higher-order quantum reservoir computing, arXiv:2006.08999 (2020).

\bibitem{Mart_nez_Pe_a_2020}
R. Martínez-Peña,  J. Nokkala, G. L. Giorgi, R. Zambrini, and M. C. Soriano, Information processing capacity of spin-based quantum reservoir computing systems, Cogn. Comput. \textbf{15}, 1440–1451 (2020).

\bibitem{Suzuki_2022}
Y. Suzuki, Q. Gao, K. Pradel, K. Yasuoka, and N. Yamamoto, Natural quantum reservoir computing for temporal information processing, Sci. Rep. \textbf{12}, 1353 (2022).  

\bibitem{Ghosh_2021}
S. Ghosh, K. Nakajima, T. Krisnanda, K. Fujii, and T. C. H. Liew, 
Quantum neuromorphic computing with reservoir computing networks, Adv. Quantum Tech. \textbf{4}, 2100053 (2021).

\bibitem{martinez2023quantum}
R. Mart{\'\i}nez-Pe{\~n}a and J.-P. Ortega, Quantum reservoir computing in finite dimensions, \pre{107}, 035306 (2023).
  
\bibitem{Bravo_2022}
R. A. Bravo, K. Najafi, X. Gao, and S. F. Yelin, Quantum reservoir computing using arrays of rydberg atoms, PRX Quantum \textbf{3}, 030325 (2022).

\bibitem{Govia_2021}
L. C. G. Govia, G. J. Ribeill, G. E. Rowlands, H. K. Krovi, and T. A. Ohki, Quantum reservoir computing with a single nonlinear oscillator, Phys. Rev. Research \textbf{3}, 013077 (2021).

\bibitem{Ghosh_2019}
S. Ghosh, A. Opala, M. Matuszewski, T. Paterek, and T. C. H. Liew, Quantum reservoir processing, npj Quantum Inf. \textbf{5}, 35 (2019).

\bibitem{Gross_2010}
D. Gross, Y.-K. Liu, S. T. Flammia, S. Becker, and J. Eisert,  Quantum state tomography via compressed sensing, \prl{105}, 150401 (2010).

\bibitem{ghosh2021realising}
S. Ghosh, T. Krisnanda, T. Paterek, and T. C. H. Liew,
Realising and compressing quantum circuits with quantum reservoir computing, Commun. Phys. \textbf{4}, 105 (2021).

\bibitem{Angelatos_2021}
G. Angelatos, S. A. Khan, H. E. T{\"u}eci, Reservoir computing approach to quantum state measurement, Phys. Rev. X \textbf{11}, 041062 (2021).

\bibitem{nakajima2018reservoir}
K. Nakajima and I. Fischer, \emph{Reservoir computing theory, physical implementations, and applications} (Springer Singapore, 2021)

\bibitem{Garc_a_Beni_2023}
J. Garc\'ia-Beni, G. L. Giorgi, M. C. Soriano, and R. Zambrini, Scalable photonic platform for real-time quantum reservoir computing, Phys. Rev. Applied, \textbf{20}, 014051 (2023).

\bibitem{Chen_2020}
J. Chen, H. I. Nurdin, and N. Yamamoto, Information processing on noisy quantum computers, Phys. Rev. Applied, \textbf{14}, 024065 (2023).

\bibitem{Agnew_2011}
M. Agnew, J. Leach, M. McLaren, F. S. Roux, and R. W. Boyd, Tomography of the quantum state of photons entangled in high dimensions, \pra{84}, 062101 (2011). 

\bibitem{Llodr__2022}
G. Llodr\'a, C. Charalambous, G. L. Giorgi, R. Zambrini, Benchmarking the role of particle statistics in quantum reservoir computing, Adv. Quantum Tech. \textbf{6}, 2200100 (2022).

\bibitem{Nokkala_2024}
J. Nokkala, G. L. Giorgi, and R. Zambrini, Retrieving past quantum features with deep hybrid classical-quantum reservoir computing, Mach. Learn. Sci. Technol. \textbf{5}, 035022 (2024).

\bibitem{Sornsaeng_2024}
A. Sornsaeng, N. Dangniam, and T. Chotibut, Quantum next generation reservoir computing: an efficient quantum algorithm for forecasting quantum dynamics, Quantum Mach. Intell. \textbf{6}, 57 (2024).

\bibitem{Mart_nez_Pe_a_2021}, 
R. Martínez-Pe\~{n}a, G. L. Giorgi, J. Nokkala, M. C. Soriano, and R. Zambrini, Dynamical Phase Transitions in Quantum Reservoir Computing, \prl{127}, 100502 (2021).

\bibitem{Dudas_2023}
J. Dudas, B. Carles, E. Plouet, F. A. Mizrahi, J. Grollier, and M. D. Markovi\'c, Quantum reservoir computing implementation on coherently coupled quantum oscillators, npj Quantum Inf. \textbf{9}, 64 (2023).

\bibitem{Li_2020}
Y. Li, R.-G. Zhou, R.-Q. Xu, J. Luo, and W. Hu, A quantum deep convolutional neural network for image recognition, Quantum Sci. Technol. \textbf{5}, 044003 (2020).

\bibitem{Peruzzo_2014}
A. Peruzzo, J. McClean, P. Shadbolt, M.-H. Yung, X.-Q. Zhou, P. J. Love, A. Aspuru-Guzik, and J. L. O’Brien, A variational eigenvalue solver on a photonic quantum processor, Nat. Commun. \textbf{5},  4213 (2014) 

\bibitem{Li_2015}
X. Li, Zhaokai, X. Liu, N. Xu, J. Du, Experimental realization of a quantum support vector machine, \prl{114}, 140504 (2015).

\bibitem{Kandala_2017}
A. Kandala, A. Mezzacapo, K. Temme, M. Takita, M. Brink, J. M. Chow, and J. M. Gambetta, Hardware-efficient variational quantum eigensolver for small molecules and quantum magnets, \nat{549}, 242–246 (2017).

\bibitem{Wang_2023}
L. Wang, Y. Sun, X. Zhang, Quantum adversarial transfer learning, Entropy, \textbf{25}, 1090 (2023).

\bibitem{Cong_2019}
I. Cong, S. Choi, and M. D. Lukin, Quantum convolutional neural network, Nat. Phys. \textbf{15}, 1273-1278 (2019).

\bibitem{Wang_2021}
L. Wang, Y. Sun, and X. Zhang, Quantum deep transfer learning, \njp{23}, 103010 (2021).

\bibitem{Biamonte_2017}
J. Biamonte, P. Wittek, N. Pancotti, P. Rebentrost, N. Wiebe, and S. Lloyd, Quantum machine learning, \nat{549}, 195-202 (2017).

\bibitem{Havl_ek_2019} 
V. Havl\'i\v cek, A. C\'orcoles, D. Antonio K. Temme, and A. W. Harrow, A. Kandala, J. M. Chow, and J. M. Gambetta, Supervised learning with quantum-enhanced feature spaces, \nat{567}, 209–212 (2019).

\bibitem{Nielsen_2012}
M. A. Nielsen and I. L. Chuang, \emph{Quantum computation and quantum information: 10th anniversary edition} (Cambridge University Press, Cambridge, 2000).

\bibitem{F_rrutter_2024}
F. F\"urrutter, G. Mu\~{n}oz-Gil, and H. J. Briegel, Quantum circuit synthesis with diffusion models, Nat. Mach. Intell. \textbf{6}, 515-524 (2024).

\bibitem{West_2023}
M. T. West, S.-L. Tsang, J. S. Low, C. D. Hill, C. Leckie, L.C. L. Hollenberg, S. M. Erfani, and M. Usman, Towards quantum enhanced adversarial robustness in machine learning, Nat. Mach. Intell. \textbf{5}, 581–589 (2023).

\bibitem{Haug_2023}
T. Haug, C. N. Self, and M. S. Kim, Quantum machine learning of large datasets using randomized measurements, Mach. Learn. Sci. Technol. \textbf{4}, 015005 (2023).

\bibitem{Huijgen_2024}
O. Huijgen, L. Coopmans, P. Najafi, M. Benedetti, and H. J. Kappen, 
Training quantum Boltzmann machines with the $\beta$-variational quantum eigensolver, Mach. Learn. Sci. Technol. \textbf{5}, 025017 (2024).

\bibitem{Senokosov_2024}
A. Senokosov, A. Sedykh, A. Sagingalieva, B. Kyriacou, A. Melnikov, Quantum machine learning for image classification, Mach. Learn. Sci. Technol. \textbf{5}, 015040 (2024).

\bibitem{Govia_2022}
L. C. G. Govia, G. J. Ribeill, G. E. Rowlands, and T. A. Ohki, Nonlinear input transformations are ubiquitous in quantum reservoir computing, Comput. Eng. \textbf{2}, 014008 (2022).

\bibitem{Domingo_2022}
L. Domingo, G. Carlo, and F. Borondo, Optimal quantum reservoir computing for the noisy intermediate-scale quantum era, \pre{106}, L043301 (2022).

\bibitem{Domingo_2024}
L. Domingo, F. Borondo, G. Scialchi, A. J. Roncaglia, G. G. Carlo, and D. A. Wisniacki, Quantum reservoir complexity by the Krylov evolution approach, \pra{110}, 022446 (2024).

\bibitem{Hu_2024}
F. Hu, S. A. Khan, N. T. Bronn, G. Angelatos, G. E. Rowlands, G. J. Ribeill, and H. E. T\"ureci, Overcoming the coherence time barrier in quantum machine learning on temporal data, Nat. Commun. \textbf{15}, 7491 (2024).

\bibitem{suzuki2022natural}
Y. Suzuki, Q. Gao, K. C. Pradel, K. Yasuoka, and N. Yamamoto, Natural quantum reservoir computing for temporal information processing, Sci. Rep. \textbf{12}, 1353 (2022).

\bibitem{Molteni_2023}
R. Molteni, C. Destri, and E. Prati, Optimization of the memory reset rate of a quantum echo-state network for time sequential tasks, \pla{465}, 128713 (2023).

\bibitem{Kubota_2023}
T. Kubota, Y. Suzuki, S. Kobayashi, Q.-H. Tran, N. Yamamoto, and K. Nakajima, Temporal information processing induced by quantum noise, Phys. Rev. Research \textbf{5}, 023057 (2023).

\bibitem{yasuda2023quantum}
T. Yasuda, Y. Suzuki, T. Kubota, K. Nakajima, Q. Gao, W. Zhang, S. Shimono, H. I. Nurdin, and N. Yamamoto, Quantum reservoir computing with repeated measurements on superconducting devices, arXiv:2310.06706 (2023).

\bibitem{Gonon_2020}
L. Gonon and J.-P. Ortega, Reservoir computing universality with stochastic inputs, IEEE Trans. Neural Netw. Learning Syst. \textbf{31}, 100-112 (2020).

\bibitem{Franz_2006}
M. O. Franz and B. Sch\"olkopf, A unifying view of Wiener and Volterra theory and polynomial kernel regression, Neural Comput. \textbf{18}, 3097-3118 (2006).

\bibitem{Paulsen_2016}
V. I. Paulsen and M. Raghupathi, \emph{An introduction to the theory of reproducing kernel Hilbert spaces} (Cambridge University Press, Cambridge, 2016).

\bibitem{Bollt_2021}
E. Bollt, On explaining the surprising success of reservoir computing forecaster of chaos? The universal machine learning dynamical system with contrast to VAR and DMD, Chaos \textbf{31}, 013131 (2021).

\bibitem{Verstraeten_2007}
D. Verstraeten, B. Schrauwen, M. D’Haene, and D. Stroobandt, An experimental unification of reservoir computing methods, Neural Netw. \textbf{20}, 391–403 (2007).

\bibitem{Maass_2002}
W. Maass, T. Natschl\"ager, and H. Markram, Real-time computing without stable states: a new framework for neural comput Based on perturbations, Neural Comput. \textbf{14}, 2531-2560 (2002).

\bibitem{devaney2018introduction}
R. Devaney, \emph{An introduction to chaotic dynamical systems} (CRC Press, Boca Raton, 2018).

\bibitem{Nambu_2002}
Y. Nambu, K. Usami, Y. Tsuda, K. Matsumoto, and K. Nakamura, Generation of polarization-entangled photon pairs in a cascade of two type-I crystals pumped by femtosecond pulses, \pra{66}, 033816 (2002).

\bibitem{Kwiat_1999}
P. G. Kwiat, E. Waks, A. G. White, I. Appelbaum, and P. H. Eberhard, Ultrabright source of polarization-entangled photons, \pra{60}, R773-R776 (1999).

\bibitem{Qiang_2018} 
X. Qiang, X. Zhou, J. Wang, C. M. Wilkes, T. Loke, S. O'Gara, L. Kling, G. D. Marshall, R. Santagati, T. C. Ralph, J. B. Wang, and J. L. O'Brien, and M. G. Thompson, and J. C. Matthews, Large-scale silicon quantum photonics implementing arbitrary two-qubit processing, Nat. Photon. \textbf{12}, 534-539 (2018).

\bibitem{Wei_2005}
T.-C. Wei, J. B. Altepeter, D. Branning, P. M. Goldbart, D. F. V. James, E. Jeffrey, P. G. Kwiat, S. Mukhopadhyay, and N. A. Peters,  Synthesizing arbitrary two-photon polarization mixed states, \pra{71}, (2005).

\bibitem{Domingo_2023}
L. Domingo, G. Carlo, F. Borondo, Taking advantage of noise in quantum reservoir computing, Sci. Rep. \textbf{13}, 8790 (2023).


\end{thebibliography}
\end{document}